\documentclass[submit]{smj}

\usepackage{caption}
\usepackage{subcaption}
\usepackage{float}
\usepackage{longtable}
\usepackage{colortbl}
\usepackage[belowskip=-25pt,aboveskip=0pt]{caption}
\usepackage{titlesec}
\usepackage{tikz}
\usetikzlibrary{shapes.geometric, arrows}
\titlespacing\section{0pt}{12pt plus 4pt minus 2pt}{0pt plus 2pt minus 2pt}
\titlespacing\subsection{0pt}{12pt plus 4pt minus 2pt}{0pt plus 2pt minus 2pt}
\titlespacing\subsubsection{0pt}{12pt plus 4pt minus 2pt}{0pt plus 2pt minus 2pt}

\Author{Muhammad Mahmudul Hasan\Affil{1, 2}, 
        Jonathan A. Cumming\Affil{1} 
}
\AuthorRunning{Hasan \& Cumming}

\Affiliations{

\item Department of Mathematical Sciences, 
      Faculty of Science, 
      Durham University,
      United Kingdom

\item Department of Statistics,
      University of Dhaka,
      Bangladesh
}  
\CorrAddress{Muhammad Mahmudul Hasan\\
             Department of Mathematical Sciences, 
             Durham University, 
             Upper mountjoy\\ 
             DH1 3LE, United Kingdom}
\CorrEmail{muhammad.m.hasan@durham.ac.uk}
\CorrPhone{+44 (0) 774 559 2896}
\CorrFax{+44 (0)191 334 3124}

\Title{A Bayesian hierarchical framework for emulating a complex crop yield simulator}
\TitleRunning{A Bayesian hierarchical framework for crop model}

\Abstract{Emulation of complex computer simulations have become an effective tool in the exploration of the behaviour of the simulated processes. Agriculture is one such area where the simulation of crop growth, nutrition, soil condition and pollution could be invaluable in any land management decisions. In this paper, we study output from the EPIC simulation model to investigate the behaviour of crop yield in response to changes in inputs such as fertilizer levels, soil, steepness, and other environmental covariates. We build a model for crop yield around a non-linear Mitscherlich Baule growth model to make inferences about the response of crop yield to changes continuous input variables (fertiliser levels), as well as exploring the impact of categorical factor inputs such as land steepness and soil type. A Bayesian hierarchical approach to the modelling was taking for mixed inputs, requiring Markov Chain Monte Carlo simulations to obtain samples from the posterior distributions, to validate and illustrate the results, and to carry out model selection. Our results highlight a strong response of yield to nitrogen, but surprisingly a weak response to phosphorus and also shows the substantial improvement of the model after adding factor effects response to maximum yield for this particular simulator configuration and catchment.
}
\Keywords{ Bayesian hierarchical; Epic model; Fertilization; MCMC; Yield model

}

\begin{document}
\maketitle
\section{Introduction}

Statistical modelling, or emulation, of the output of computer simulation has become an increasingly useful tool for the analysis of complex systems within the sciences \cite{cumming2010bayes, jackson2020understanding, hasan2021bayes}. While a computer simulation can be constructed to capture our best understanding of the mathematical and scientific processes within the system, we are typically left with substantial uncertainties on the precise operation of the system. These uncertainties can range from simple uncertainties on the values of the parameters, to more epistemic uncertainties surrounding our understanding of the science represented by the simulation \cite{goldstein2009reified}. Consequently, this motivates a statistical treatment of the analysis of such data and models.

Computer simulations are widely used across multiple disciplines, and in this paper we focus on an application within agriculture - specifically the simulation and modelling of crop yield. The behaviour of the yield of a particular crop in response to endogenous variables such as fertilisation levels and exogenous variables such as weather has been extensively studied in the literature \cite{llewelyn1997comparison, alivelu2003comparison,paine2012fit,salo2016comparing}. Typically, such models comprise a non-linear relationship between the observed yield and the fertiliser levels for a particular crop. The model is also expected to respect a number of intuitive features of the relationship between these quantities, such as the yield should increase monotonically in response to additional fertiliser and the presence of a plateau effect beyond which additional fertilisation will have no further benefit. A particular challenge that has not been well addressed is how to adapt such models to account discrete-valued inputs, such as land characteristics or management scenarios. For our analysis, we will use the output of the EPIC (Environmental Policy Integrated Climate) simulator \cite{williams1984epic, jones1991epic}, which simulates time series of crop yield incorporating fertilizers, soil, steepness, and weather. 

The remainder of this article is organized as follows. Section 2 reviews the relevant features of the EPIC simulator, sample data output, and reviews suitable models for crop yield. In Section 3, we present our hierarchical Bayesian crop yield model. Section 4 briefly considers how to introduce discrete factor inputs into the model of Section 3. In Section 5, we present the results of the Bayesian hierarchical model fitting with model comparison and diagnostics, and Section 6 concludes the article with discussion.

\section{Simulation and Modelling of Crop Yield}
\subsection{Simulating Yield}
Our analysis is based on the Environmental Policy Integrated Climate(EPIC) model, which is a simulator of crop yield corresponding to inputs nitrogen and phosphorous, emissions of various pollutants, and also the effects of land and soil management systems  \cite{williams1984epic, jones1991epic}. The data for our analysis comprises a large-scale simulation from the EPIC simulator over a fully-factorial design. This generated a large ($\gg 20$ Tb) amount of time series output data, which was post-processed for the purposes of this study to extract annual crop yields for each design point over the sixty year period of simulation.

The key variables of interest in our investigation are
\begin{itemize}
    \item Fertiliser levels: Nitrogen ($N$) and Phosphorus ($P$). Continuous variables each specified at a grid of 13 values over $[0,100]$, where $100$ represents a pre-defined maximum level.
    \item Land characteristics: Soil type ($So$) and Steepness ($St$). Categorical variables representing terrain features not captured numerically, with five and four levels respectively.
\end{itemize}

Additional inputs representing weather and related variables were also supplied to the simulator, however we consider the available set of historical weather (rainfall, temperature etc) scenario which is offset and rotated to allow all crops to experience a range of different weather sequences. The crop which face more rotations, they have more levels of weather factor. The simulations were  performed for 24 number of different crop rotations involving 13 distinct crops. This yielded simulated crop yield output for 98 different `unique' crops in response to the key variables mentioned above such that the maize when planted after turnips and maize following potatoes, and the corresponding yield information was extracted from the same point in the simulations. We will focus on three crops from the EPIC simulator output: Spring Barley, Winter Barley, and Silage. 

\begin{figure}
     \centering
     \includegraphics[width=13cm,height=6cm]{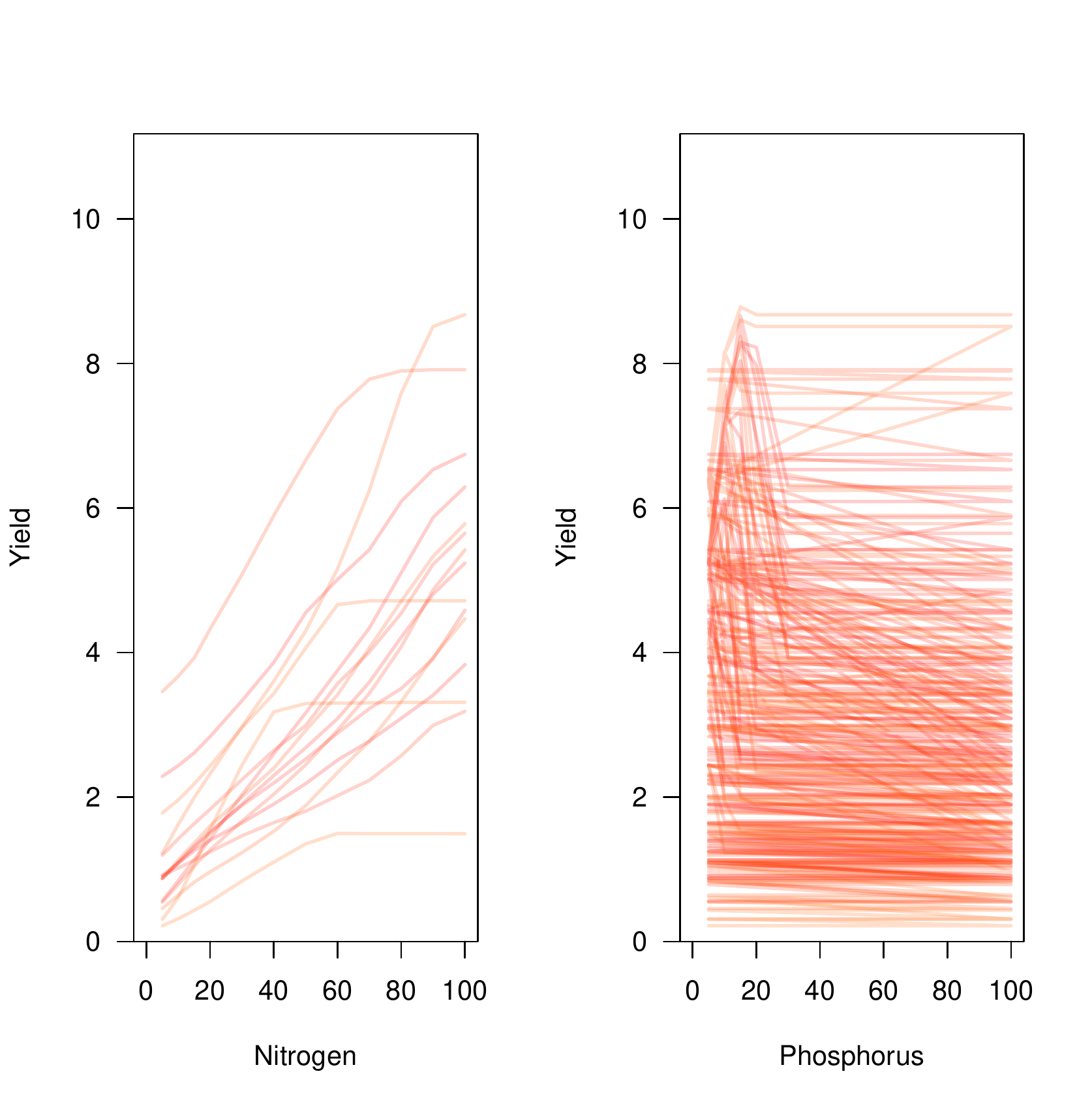}
\caption{Plots of simulated yield for a sample of 15 simulations in response to nitrogen (left panel) and phosphorus (right panel) for Spring Barley }  
\label{fig:1}
 \end{figure}

\begin{figure}
     \centering
     \includegraphics[width=13cm,height=6cm]{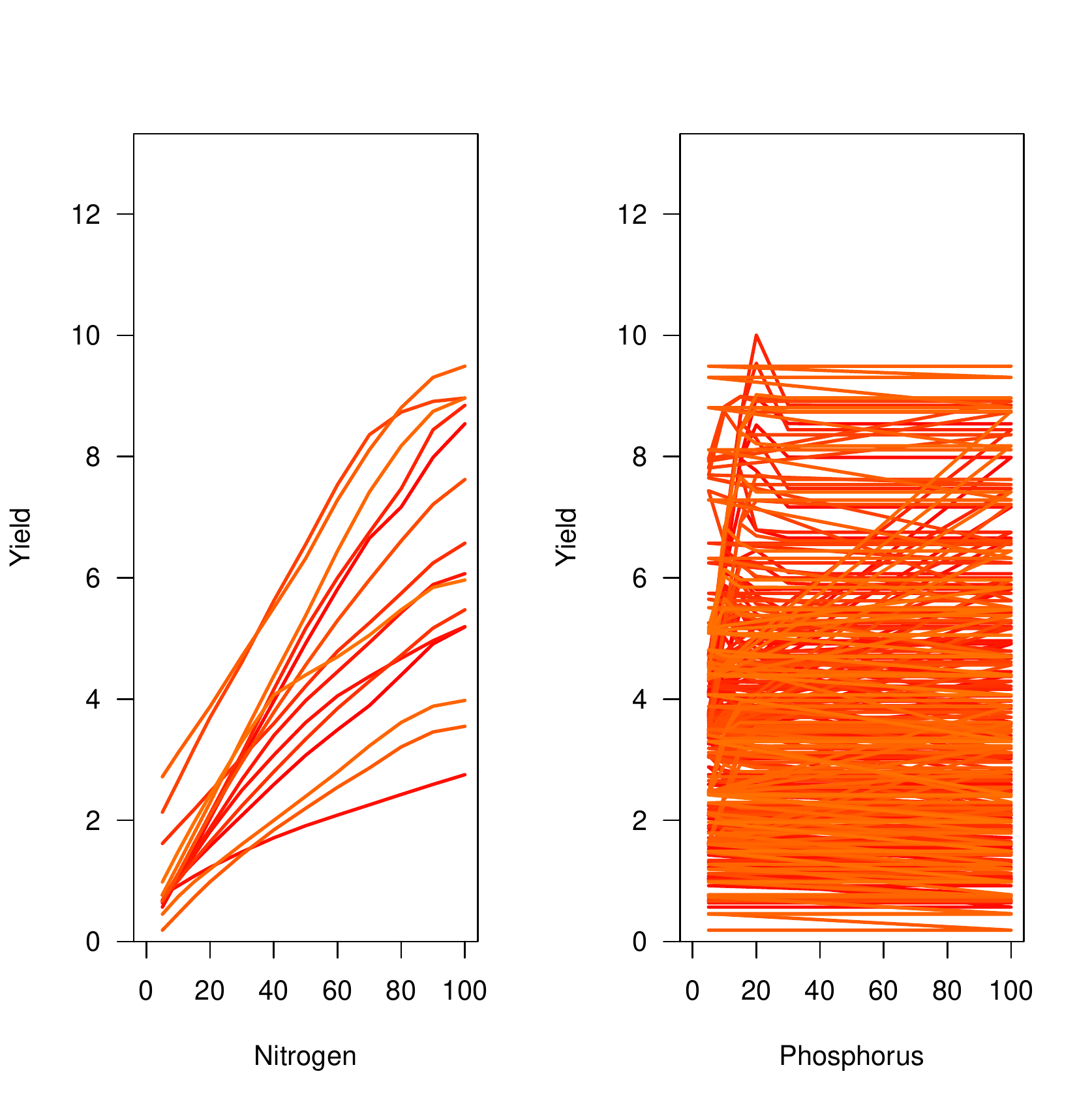}
   \caption{Plots of simulated yield for a sample of 15 simulations in response to nitrogen (left panel) and phosphorus (right panel) for Winter  Barley }   
     \label{fig:2}   
 \end{figure}

\begin{figure}
     \centering
     \includegraphics[width=13cm,height=6cm]{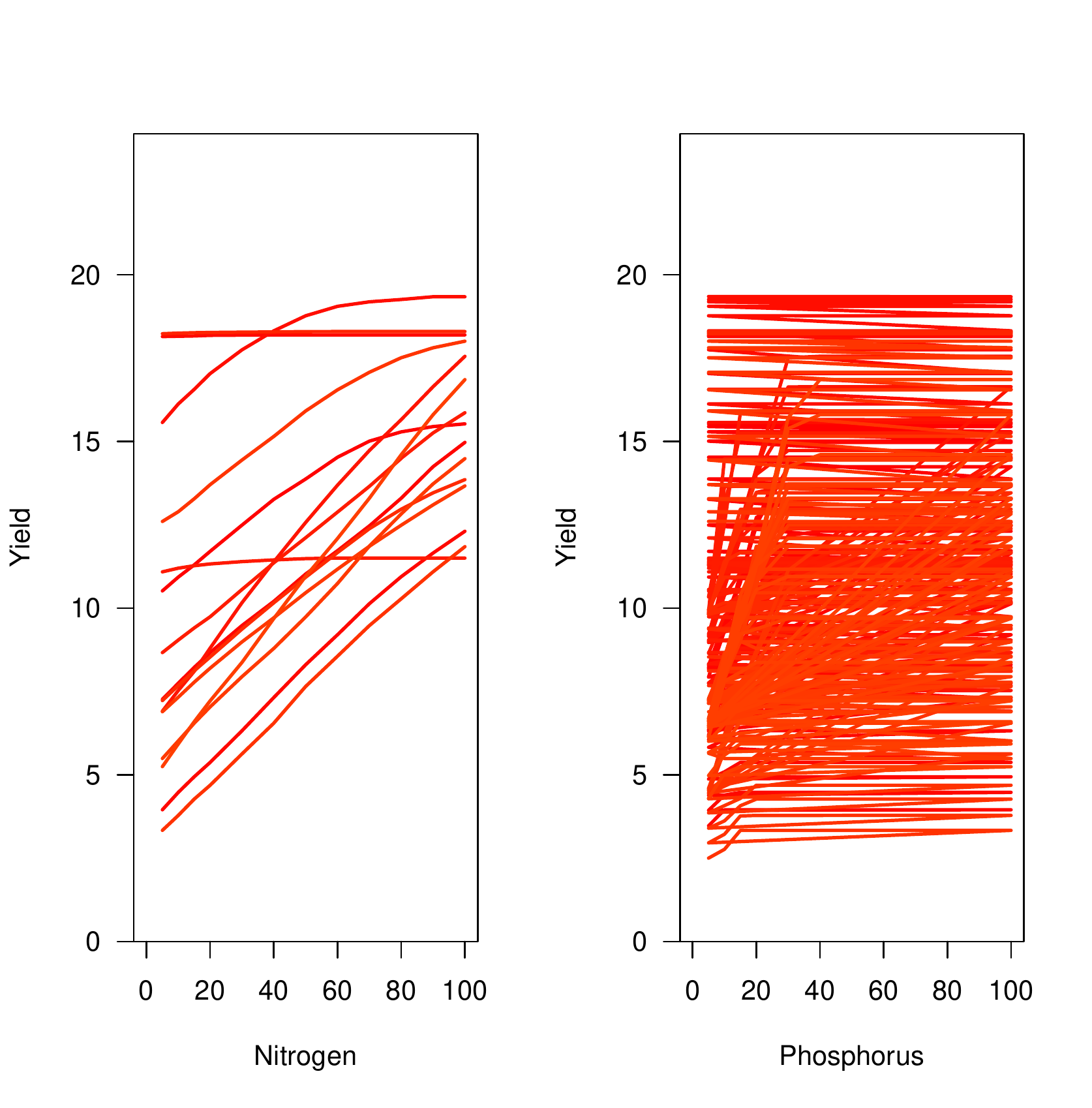}
  \caption{Plots of simulated yield for a sample of 15 simulations in response to nitrogen (left panel) and phosphorus (right panel) for Silage }  
       \label{fig:3}
 \end{figure}

Initially, in  Figure ~\ref{fig:1}, ~\ref{fig:2}, and ~\ref{fig:3} we plot the raw simulator output as yield  versus the continuous fertiliser input nitrogen and phosphorus for a subset of $15$ different combinations for three crops in particular - Spring Barley, Winter Barley, and Silage. 
We avoided to assess all combinations of greater than $1000$ because of much data makes it hard to read due to over-plotting, failure to clear variation among the simulations, and difficult to get a clear trend from all the unique combinations. We also excluded the level $0$ from the both inputs due to its unexpected non-physical behavior. From Figure ~\ref{fig:1}, ~\ref{fig:2}, and ~\ref{fig:3} it can be said that yield shows monotonic increasing trend, and plateau for all crops but some flat trends for a few of the unique combinations for silage in response to input nitrogen. On the other hand, clear flat response and some unusual trend with respect to phosphorous for all of the explored three crops.
 
%%\vspace{-20pt}
\subsection{Modelling Crop Yield}
%%\vspace{-20pt}
The response of crop yield to inputs such as fertilisation has been extensively studied in the literature
\cite{frank1990comparison,grimm1987liebig,llewelyn1997comparison} providing many useful insights into the expected or desirable behaviour of any yield response model. Some of the key features of a crop yield model include;
%%\vspace{-20pt}
\begin{enumerate}
    \item Monotonicity: crop yield should be monotonically increasing (or at least non-decreasing) in response to increases in fertiliser levels.
    \item Growth plateau: crop yield will increase, but ultimately plateau in response to increase fertiliser levels to represent the saturation of the plant and diminishing returns of yield with additional fertilizer.
    \item Input substitution: is the replacement of one input to other for a certain crop \cite{sheng2016input}. So an increase in $N$ can be substituted for an increase in $P$ and vice versa, rather than affecting the crop yield in different and non-exchangeable ways. 
    \item Returns to scale: is used to measure the proportional change of the yield with the change of inputs $N$ and $P$.
\end{enumerate}

Various functional forms of yield, $Y=(Y_1,Y_2,,,Y_n)$, in response to Nitrogen, $N=(N_1,N_2,...,N_n)$, and phosphorous, $P=(P_1,P_2,...,P_n)$, have been proposed in the literature, and can be loosely categorised into three groups and we summarise these yield models in Table \ref{t:YieldModels}:
%%\vspace{-20pt}
\begin{enumerate}
    \item \textbf{Linear Models:} Multiple Regression, Quadratic, and Square Root function.
    \item \textbf{Nonlinear Models:} Power function, Gompertz function, Logistic function, and Mitscherlich-Baule function.
    \item \textbf{Threshold Models:} Linear Von Liebig function, and Nonlinear Von Liebig function.
\end{enumerate}

\small
\begin{table}[H]
\caption{Summary of models of yield, $Y$, in response to nitrogen, $N$, and phosphorous, $P$.}
\label{t:YieldModels}
\begin{center}
\setlength\tabcolsep{5pt}
\begin{tabular} { |p{4cm}|l|p{4cm}| }  
\hline
Name & Function & Features \\
\hline
Linear & 
$\begin{array}{lcl}
Y
&=& \beta_0+\beta_1N+\beta_2P
\end{array}$ 
&Linearity\\ 

Quadratic & 
$\begin{array}{lcl}
Y &=& \beta_0+\beta_1N+\beta_2P\\
&& +\beta_3N^2+\beta_4P^2+\beta_5NP
\end{array}$ 
& Input substitution \\ 

Square Root & 
$\begin{array}{lcl}
Y &=& \beta_0+\beta_1N+\beta_2P\\
&& +\beta_4(P)^{1/2}+\beta_5(NP)^{1/2}
\end{array}$ 
& Input substitution, no plateau\\

Power Model & 
$\begin{array}{lcl}
Y &=& \beta_0N^{\beta_1}P^{\beta_2}
\end{array}$ 
& returns to scale\\

Gompertz &  
$\begin{array}{lcl}
Y &=& \beta_0 \exp(-\beta_1 e^{-\beta_2N})
\end{array}$
& plateau\\

Logistic & 
$\begin{array}{lcl}
Y &=& \beta_0/(1+\beta_1 e^{-\beta_2N})
\end{array}$
& plateau\\

Mitscherlich-Baule & 
$\begin{array}{lcl}
Y &=& \beta_0[1-\exp(-\beta_1 (\beta_2-\beta_3 N))] \\
&& \times [1-\exp(-\beta_4 (\beta_5-\beta_6 P))]
\end{array}$
& monotonicity\\

Linear Von Liebig &
$\begin{array}{lcl}
Y &=& \min[\beta_0, \beta_1+\beta_3N, \beta_2+\beta_4P]
\end{array}$
& plateau\\

Nonlinear Von Liebig &
$\begin{array}{lcl}
Y
&=& \min[\beta_0 (1-\beta_1 \exp((-\beta_2 N))), \\
&& ~~~~~~\beta_0 (1-\beta_3 \exp((-\beta_4 \times P)))]
\end{array}$
&plateau\\
\hline
\end{tabular}
\end{center}
\end{table}

\begin{figure}
     \centering
     \begin{subfigure}[b]{0.32\textwidth}
         \centering
         \includegraphics[width=\textwidth]{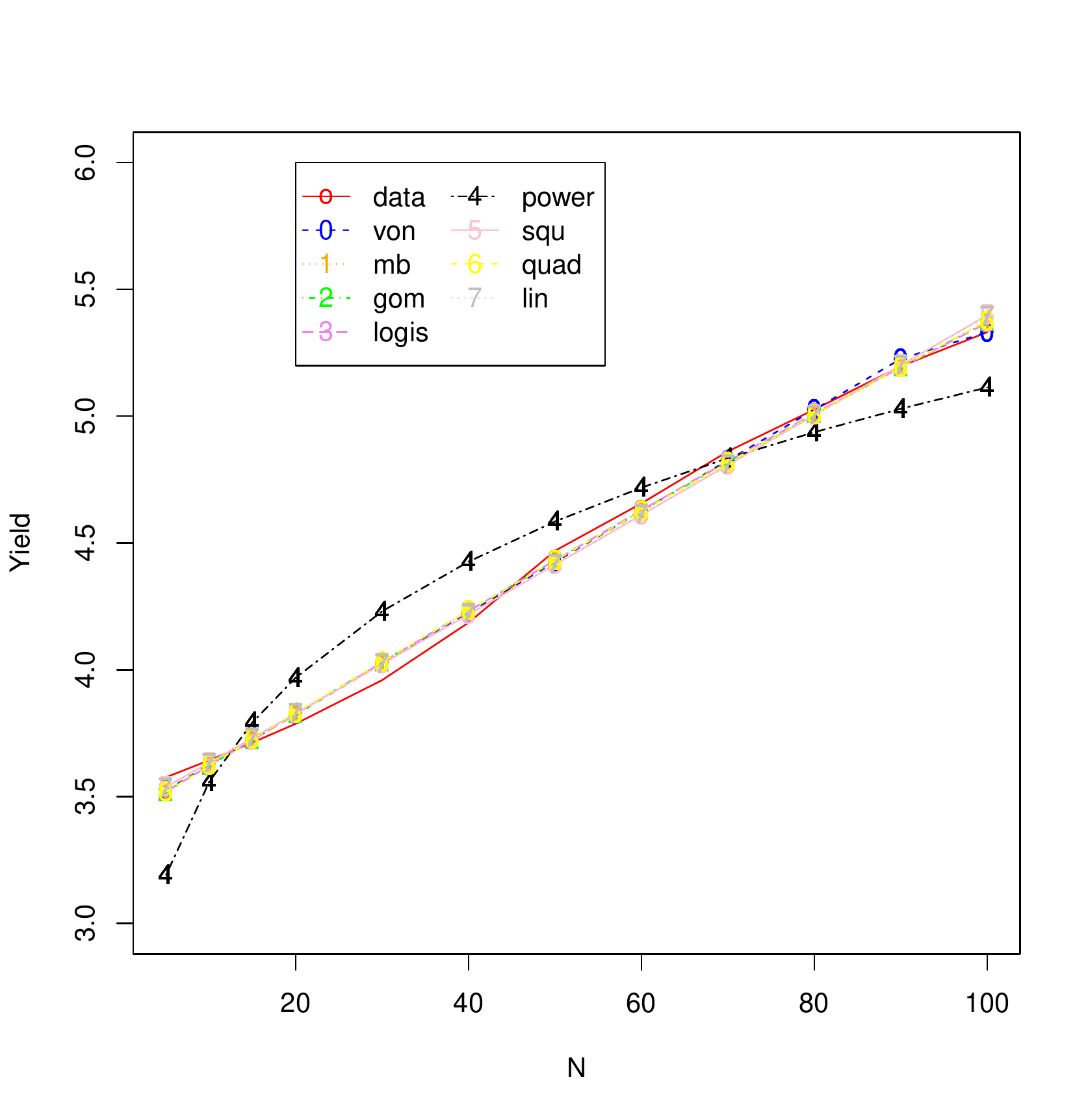}
     \end{subfigure}
     \hfill
     \begin{subfigure}[b]{0.32\textwidth}
         \centering
         \includegraphics[width=\textwidth]{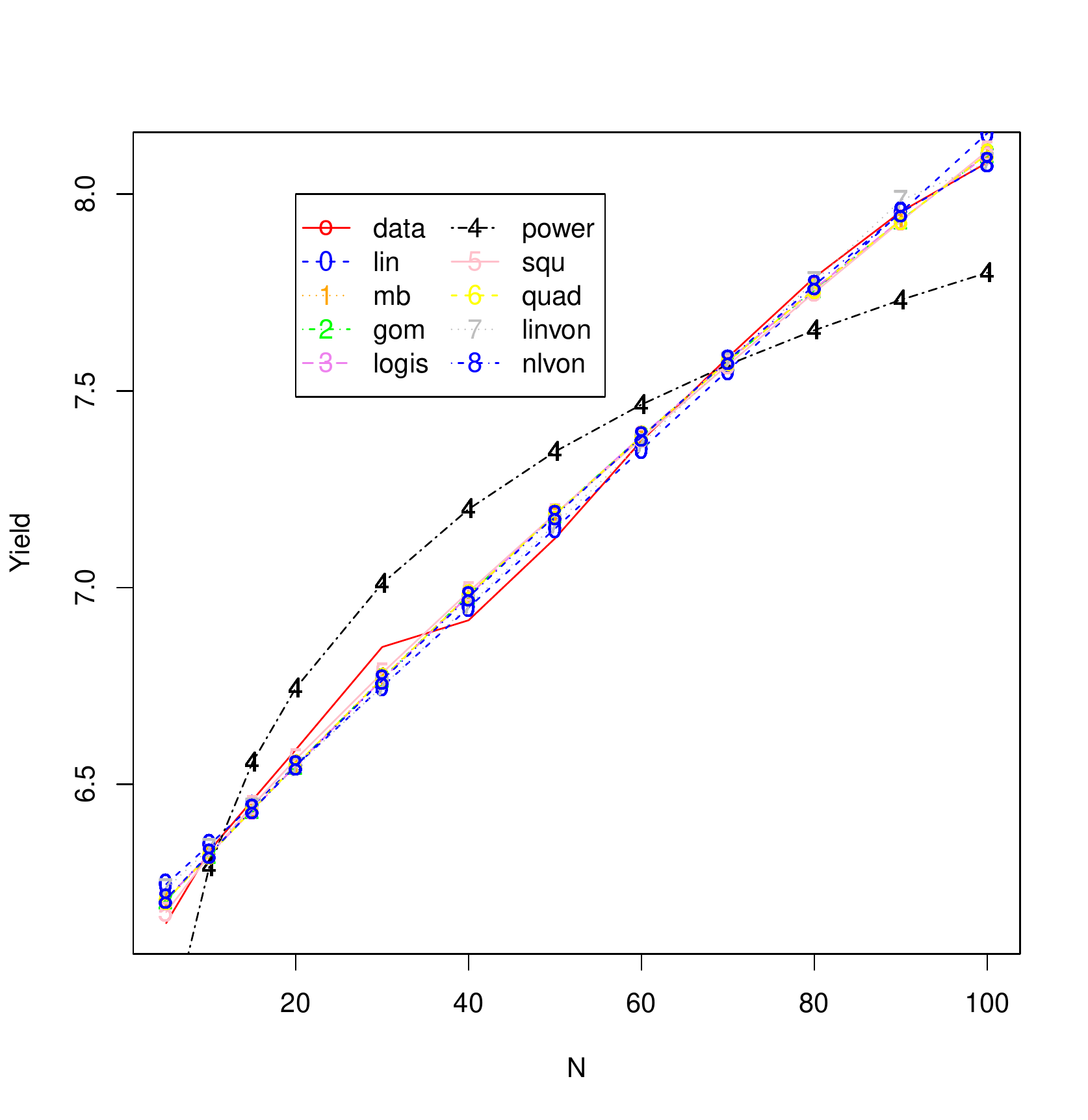}
     \end{subfigure}
      \hfill
     \begin{subfigure}[b]{0.32\textwidth}
         \centering
         \includegraphics[width=\textwidth]{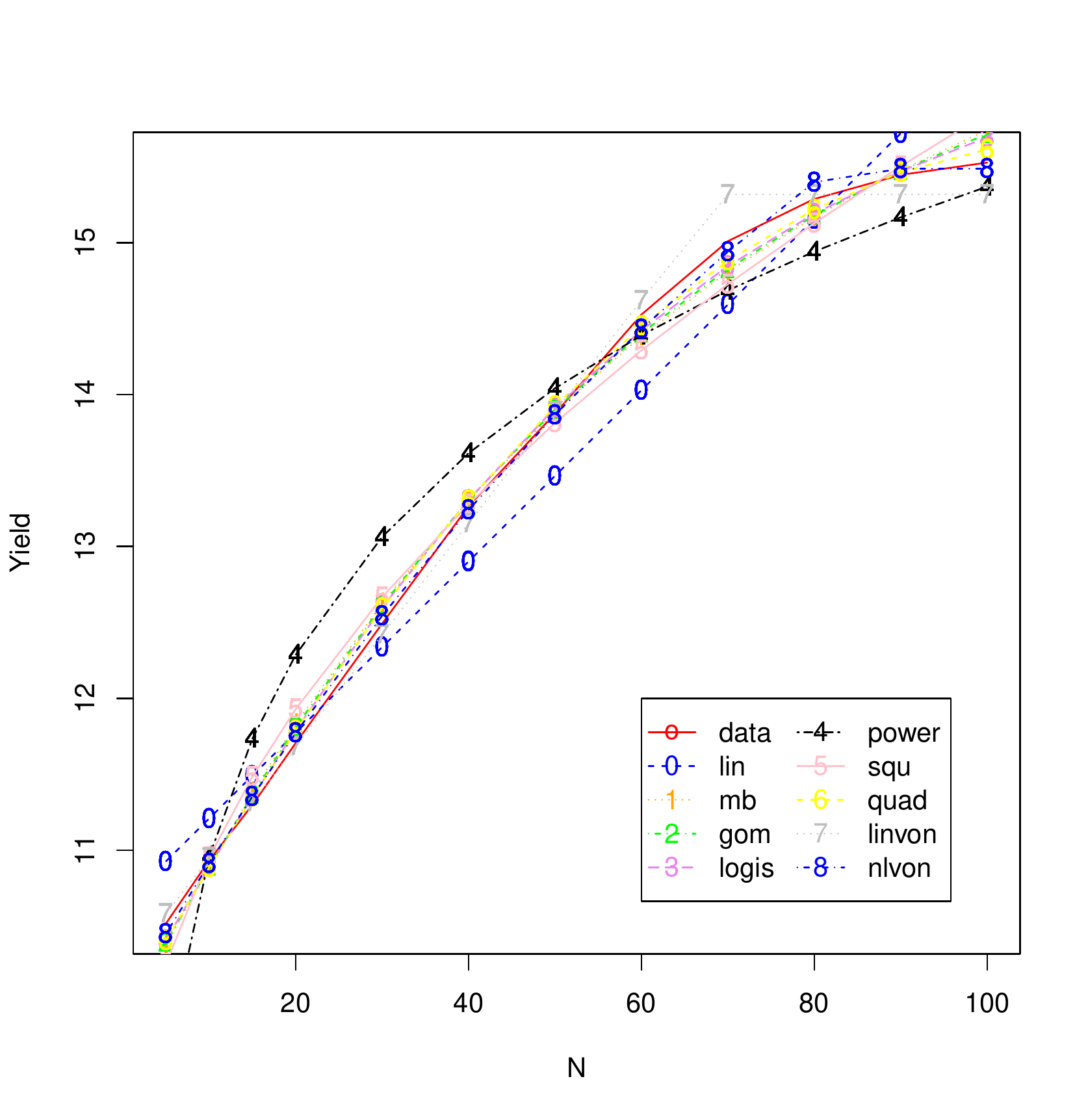}
     \end{subfigure}
   \caption{Plot of the fitted yield response models for the spring barley (left panel); winter barley (middle panel); silage (right panel)}      
 \label{fig:4}    
\end{figure}

For our model-fitting, we have explored the above nine different yield response models for both nitrogen and phosphorus. We have compared these models to the data (red line) in Figure ~\ref{fig:4} for Spring Barley, Winter Barley, and Silage crops respectively. From the Figure  ~\ref{fig:4}, we can see that all models are able to achieve reasonably close fits to the data, apart from the power model for the Spring Barley and Winter Barley data. Silage is a bit different from the Spring Barley and Winter Barley, where we can see more spread and variation between the different fitted models with a visible plateau for Von Liebig's models but a higher  residual standard error (RSE) of $0.921$. We see no input substitution in our results since $P$ has no effect, which make the use of linear models infeasible and no visible plateau for the Gompertz as well as Logistic. The Mitscherlich-Baule model shows monotonic increasing trend for all crops and it is expected that yield should increase with fertilizer levels and respond to one, other, or both fertilizers depending on the crop attributes. Due to the strong similarity between the fitted yield responses; only model to achieve the attribute of crop yield for all crops; smaller residual standard error (RSE) of $0.08$ and also the preference in the literature \cite{llewelyn1997comparison,finger2008application} over others, we proceed only with the MB yield response function for the remainder of our analysis.

\section{Bayesian Uncertainty Analysis}
%%\vspace{-20pt}
The ultimate goal for this section is to set up a Bayesian framework for both quantitative and qualitative inputs based on the MB model mean function, draw posterior samples by using MCMC algorithms, and validate the model using different diagnostic tools for the selected MCMC algorithm.

While the classical approach to modeling is centred on using likelihood for both observed and unobserved data, the Bayesian approach combines prior beliefs with the likelihood to obtain the posterior probability distribution. The Bayesian approach draws this posterior distribution via Bayes rule
\begin{equation} 
 \pi(\theta|y)=\frac{\pi(y|\theta) \pi(\theta)}{\pi(y)},
\end{equation}
where $\pi(\theta|y)$ is posterior distribution of the parameters $\theta$ after gathering the data, $\pi(y|\theta)$ is the likelihood and quantifies the information from observing the data, $\pi(\theta)$ is prior distribution of $\theta$ representing the information about the $\theta$ is gathered from observed data, and $\pi(y)$ is the marginal distribution of the data.\
Following \cite{blasco2003bayesian} we can also consider our non-linear hierarchical Bayesian framework for quantitative inputs based on the following three stages: 
%%\vspace{-20pt}
\begin{enumerate}
    \item Stage I: Bayesian Modelling for continuous covariates.
    \item Stage II: Posterior Samples from MCMC.
    \item Stage III: Model comparison and validation.
\end{enumerate} 
After then, we can adopt the mean function for the factor effect similar to the mentioned stages. The layout for the Bayesian hierarchical framework is described as follows,
%%\vspace{-20pt}
\subsection{Stage I: Bayesian Modelling for continuous inputs}\label{3.1}
%%\vspace{-20pt}
For each unique crop, we assume the yield $Y_i$; $i=1, 2,...,n$ follow the normal distribution with mean $\mu(N_i,P_i| \mathcal{B})$ and variance $\sigma$.
 \begin{equation}
         Y_i \sim N(\mu(N_i,P_i| \mathcal{B}),\sigma)
      \end{equation}
Initially, we consider modelling yield as a function of the continuous covariates nitrogen (N) and phosphorus (P) for the study and the Mitscherlich-Baule model as mean response function. Thus, the MB mean yield function from the Table \ref{t:YieldModels} can be parameterized in the following way,
 \begin{equation}
      \mu(N_i,P_i| \mathcal{B})=\beta_0[1-\exp(-\beta_1-\beta_2 N_i)][1-\exp(-\beta_3-\beta_4 P_i)],
 \end{equation}
where parameter $\beta_0>0$ is the maximum yield, $\beta_1,\beta_2$ are the intercept and slope for the nitrogen input, and $\beta_3,\beta_4$ are the intercept and slope for the phosphorus input.

Since the parameter $\beta_0$ of our model is the maximum yield, which is always be positive. The parameters $\beta_1,\beta_2$ for the nitrogen input, $\beta_3,\beta_4$ for the phosphorus input provided the positive estimates from the nonlinear modeling and also showed the increasing positive trend at the graphical analysis for the selected three crops. Due to the positivity nature of the MB model \cite{rosenzweig1999wheat} and estimates from our crop yield modelling, we have considered a gamma distribution as a possible prior for all of our five parameters of the Mistcherlich-Baule model. Following similar arguments, we have also considered the gamma prior for our error variance $\sigma$. 
 Thus
 \begin{equation}
 \begin{aligned}
     \beta_j &\sim Gamma(\alpha_j,\lambda_j);  j=0,1,2,3,4\\
     \sigma &\sim Gamma (u,v) 
     \end{aligned}
 \end{equation}

The general form of our joint prior distribution of all parameters and error term $\sigma$ can be presented as follows,
\begin{equation}
    \pi(\beta_i; \sigma)
 \propto{(\sigma)^{u-1}\exp(-\sigma v)\exp(-\sum_{j=0}^{4}{\beta_j\lambda_j})\displaystyle\prod_{j=0}^{4}\beta_j^{\alpha_j{-1}}}
\end{equation} 
where $\alpha_j$ and $\lambda_j$ are the shape parameters and rate parameters for the $\beta_j$; $u$ and $v$ are the shape parameters and rate parameters for the variance $\sigma$.

 Considering the yield observations $Y_i$ as a vector $\mathbf{Y}=(Y_1,Y_2,...,Y_n)$ and vector $\mathcal{B}=(\beta_0, \beta_1, \beta_2, \beta_3, \beta_4)$; assuming conditional independence given $\mathcal{B}$, $\sigma$ we can write the likelihood function; 
 %%\vspace{-10pt}
 \begin{equation}
       L(\mathbf{Y}|\mathcal{B},\sigma)=\displaystyle\prod_{i=1}^{n}\pi(\mathbf{Y}|\mathcal{B},\sigma)
 \end{equation}
 We have already assumed that our yield model follows a normal distribution $(3.2)$, so the likelihood function can be written as,
%%\vspace{-10pt}
 \begin{equation}
 \begin{aligned}
       L(\mathbf{Y}|\mathcal{B},\sigma)
     &=\displaystyle\prod_{i=1}^{n}\frac{1}{\sqrt{2\sigma \pi}} \exp\Big[-\frac{1}{2\sigma}(Y_i-\mu_i)^2\Big]  \\
     &\propto{\sigma^{-\frac{n}{2}}\exp\Big[{-\frac{1}{2\sigma}}\sum_{i=0}^{n}(Y_i-\mu_i)^2}\Big]
 \end{aligned}
\end{equation}
Using the likelihood for an MB model for mean yield in $(3.7)$ and prior distribution in $(3.5)$, we can write the form of our posterior distribution as,\
\begin{equation}
       \pi(\mathcal{B},\sigma|\mathbf{Y})\\
       %&\propto{W^{-\frac{n}{2}}\exp[{-\frac{1}{2W}}\sum_{i=0}^{n}(Y_i-\mu_i)^2}] * (W)^{u-1}\exp(-Wv)\exp(-\sum_{i=0}^{4}{\beta_i\lambda_i})\displaystyle\prod_{i=0}^{4}\beta_i^{\alpha_i{-1}}&\\
       %&\propto{W^{u-1-\frac{n}{2}}\exp[{-\frac{1}{2W}}\sum_{i=0}^{n}(Y_i-\mu_i)^2}] * \exp(-Wv)\exp(-\sum_{i=0}^{4}{\beta_i\lambda_i})\displaystyle\prod_{i=0}^{4}\beta_i^{\alpha_i{-1}}&\\
       \propto{\sigma^{u-1-\frac{n}{2}}\exp\Big[{-\frac{1}{2\sigma}}\sum_{i=1}^{n}(Y_i-\mu_i)^2}-\sigma v-\sum_{j=0}^{4}{\beta_j\lambda_j}\Big]\displaystyle\prod_{j=0}^{4}\beta_j^{\alpha_j{-1}}
\end{equation}
 Clearly, the equation $(3.8)$ is not recognisable as a standard distribution. Thus we require the use of MCMC for drawing posterior samples.
%\vspace{-30pt}
\subsection{Stage II: Posterior Sampling from MCMC}\label{3.2}
%%\vspace{-20pt}
In the Bayesian setup, we use MCMC to generate samples from the set up of posterior probability distribution (3.8) using Hamiltonian Monte Carlo within No-U-Turn sampler. There are different set of algorithms used to generate MCMC samples on the basis of posterior distribution.

The Metropolis Hastings (MH) algorithm \cite{hastings1970monte} which generate candidate based on the full joint density distribution. This algorithm works well if the proposal distribution matches with the target distribution. For our layout choosing the appropriate initial value is very difficult and also the low acceptance, slow mixing make the MH algorithm infeasible. Gibbs sampling \cite{gelfand2000gibbs} is based on the conditional distribution and for the nonlinear mean function, it is inefficient to compute the conditional distribution for each parameters and the correlated nature among the parameters of the Mitscherlich-Baule model also makes the use of Gibbs sampling impractical for our particular hierarchical Bayesian setup.

The Hamiltonian Monte Carlo(HMC) \cite{neal2011mcmc} is becoming popular MCMC method due to its dynamical behaviour such that use differential equation system to produce marginal variance discarding the random walk. HMC calculate the gradient of the posterior distribution and it can also simulate samples with high acceptance rate and less iterations for convergence over the wider ranges of parameter space but sometimes poor selection of the parameter value drastically reduce the effectiveness of MCMC. For this nature, Hoffman and Gelman  \cite{hoffman2014no} introduced No-U-Turn sampler within HMC by the recursive algorithm for the efficiency of HMC and implemented by the Stan language used in R and SAS language \cite{team2016rstan}. It use auto differentiation to get derivatives of every parameters.

HMC efficiency strongly relies on the tuning parameters of momentum covariance, step size and the the steps number corresponding to number of iterations \cite{hoffman2014no}. So the introduction of NUTS within HMC adaptively tunes the parameters at the time of warming period and adjust the steps size number while iteration process. 

\subsection{Stage III: Model selection and validation}\label{3.3}

It is expected that not all crops will respond to both inputs, and so we use model selection tools to identify the appropriate model in-terms of the fertilizer response. We considered three possible models for yield response: (i) nitrogen only, (ii) phosphorus only, and (iii) both model, giving the following mean functions.
\begin{align}
     \mu_i^{N}&=\beta_0[1-\exp(-\beta_1-\beta_2 N_i)] \nonumber\\
    \mu_i^{P}&=\beta_0[1-\exp(-\beta_3-\beta_4 P_i)]\\
    \mu_i^{NP}&=\beta_0[1-\exp(-\beta_1-\beta_2 N_i)][1-\exp(-\beta_3-\beta_4 P_i)] \nonumber
\end{align}

Under each of the above three possible mean function models, we applied the Hamiltonian Monte Carlo within NUTS to simulate the posterior samples. We then compared the model performance of the three models via multiple model selection criteria and therefore we considered the Expected Log point wise Predictive Density(ELPD) \cite{gelman2013bayesian}, Leave-One-Out cross-validation criterion (LOOIC) \cite{vehtari2017practical}, Widely Applicable Information Criterion (WAIC) \cite{vehtari2017practical}, and the Bayes factor (BF) \cite{jeffreys1998theory}. \
Let us consider for our yield $Y_i=(Y_1,Y_2,...,Y_n)$ given parameters $\mathcal{B}$'s the prior distribution is $\pi(\mathcal{B})$ and the posterior distribution is $\pi(\mathcal{B}|Y_i)$. So the posterior predictive density (\cite{vehtari2017practical}) can be expressed as ;
\begin{equation}
    ELPD=\displaystyle\sum_{i=0}^{n}\int\pi({Y_i}|\mathcal{B}) \log \pi (\mathcal{B}| {Y_i}) d\pi
\end{equation}
where ${Y_i}$ is the yield for model fitting and can be calculated using LOOIC and WAIC. The maximum value of ELPD is considered as the best selection criterion.
The log score used for determining the predictive density can be written as;
\begin{equation}
     LPD=\displaystyle\sum_{i=0}^{n}\int\pi({Y_i}|\mathcal{B}) \log \pi(\mathcal{B}|{Y_i}) d\mathcal{B} 
\end{equation}

So in practical world the log point wise density (LPD) can be calculated from the $\pi_{posterior}(\mathcal{B})$ posterior distribution samples and the simulations can be expressed as $\mathcal{B}^k; k=1,...,K$. This is needed to apply for cross validation and to calculate WAIC.
\begin{equation}
    \widehat{LPD}=\displaystyle\sum_{i=0}^{n} \log (\frac{1}{K}\displaystyle\sum_{k=1}^{K}\pi(Y_{i}|\mathcal{B}^{k}))
\end{equation}
The LOO estimate for the Bayesian model selection \cite{vehtari2017practical} can be expressed as;
\begin{equation}
    \widehat{ELPD}_{LOO}=\displaystyle\sum_{i=0}^{n} \log \pi(Y_{i}|Y_{-i})
\end{equation}
where $\pi(Y_i|Y_{-i})$ is the LOOIC density for given yields without the $i$th yields. The minimum value of LOOIC is considered as the best selection criterion.
WAIC (\cite{watanabe2010asymptotic}) is another approach for estimating expected log point wise predictive density can be written as,
\begin{equation}
    \widehat{ELPD}_{WAIC}=\widehat{LPD}-\widehat\pi_{WAIC}
\end{equation}
 where $\widehat\pi_{WAIC}$ can be calculated from $\displaystyle\sum_{i=0}^{n} Var_{posterior}(\log \pi( \mathbf{}{Y}| \mathcal{B}))$. The minimum value of WAIC is considered as the best selection criterion.
The Bayes factor is the relative probability of two models for observed data points such that the ratio of alternative hypothesis and null hypothesis given data points can be expressed as;
\begin{equation}
    \beta_{10}=\frac{\pi (\mathbf{Y_{i}}|H_1)}{\pi(\mathbf{Y_{i}}|H_0)}
\end{equation}
The interpretation of the Bayes factor was proposed by Harold Jeffrey's in 1961 \cite{jeffreys1998theory} and then modified by Lee and Wagenmakers in 2013 \cite{lee2014bayesian} such that values less than 1 have moderate to strong evidence for null and vice versa. There has no evidence for the ratio given the value zero. 
\section {Incorporating a factor effect}\label{sec:4}
The structure of MB model itself is difficult to formulate and introducing as the mean function for fully Bayesian framework of mixed inputs is also computationally infeasible. There are number of ways to formulate the problem and we use the pragmatic approach to keep the problem simple. We first formulate the set up for continuous responses and then used the selected model to introduce factor inputs as a design matrix. 

Assume factors only influence maximum yield $\beta_0$, by shifting the value. Assuming this implication of $\beta_0$, we use a simple hierarchical structure of the following form;
\begin{equation}
    \beta_{0}=\gamma_0+X\mathbf{\gamma_{1,i}};
\end{equation}
where $X$ is the model matrix of the factor variables, and $\mathbf{\gamma_{1,i}}$ is the vector of factor effect parameters' for $i=1,2,...,k$; k is the length of the factor level and $\gamma_0$ is the coefficient similar to the maximum yield of $\beta_0$. Considering above two assumptions, the Equation (3.9) for the $N$ input now can be written as;
    %%\vspace{-20pt}
\begin{equation}
    \mu_i^{N}=(\gamma_0+X\mathbf{\gamma_{1,i}})[1-\exp(-\beta_1-\beta_2 N_i)]
\end{equation}
The priors for $\gamma_0$, $\beta_1$ and $\beta_2$ are similar to the priors of $\beta_0, \beta_1, \beta_2$ of Equation (3.4). The factor effect $\mathbf{\gamma_{1,i}}$ in Equation (4.1) can be positive or negative, so we consider the standard normal prior for our factor effects.
 Using the Equation (4.1) to Equations (3.7 - 3.8) make us possible to calculate likelihood distribution and hence posterior distribution for the factor effect. The form of posterior distribution is similar to continuous variable modelling and needed to use using HMC-NUTS MCMC algorithm to generate posterior samples due to unrecognised function.

\section{Results of Bayesian Analysis}
In this section we show and discuss the results of Bayesian hierarchical analysis corresponding to four stages mentioned in the Section 3 and Section 4. 

%\subsection{Stage 1: Bayesian Modelling for continuous covariates}
The Mistcherlich-Baule response function, upon which we base our model, has several characteristics which can be incorporated into the prior distributions for its parameters \citep{finger2008application}. 
Namely, we know that: (i) $\beta_0$ is the maximum yield which should be positive and finite; (ii) the intercept, $\beta_1$, and slope, $\beta_2$, for the nitrogen (N) response should have positive parameters; (iii) we expect similar properties from the phosphorus (P) parameters $\beta_3$, and $\beta_4$. As all parameters are required to be positive, we adopt weakly informative gamma priors to enforce this positivity. Specifically, we used the $\beta_0\sim Ga(Y_{max},1)$ for maximum yield parameter where $Y_{max}$ is the maximum observed yield value for that crop. We have used the $10\%$ of the data to assess the estimate for the coefficients using nonlinear MB model fitting as the basis for the weakly informative priors and fitted the model with the rest $90\%$ of the data.  For the intercept and slope priors corresponding to $N$, we used $\beta_1\sim Ga(3.43, 1)$ and $\beta_2\sim Ga(1, 20)$, and for $P$ we have $\beta_3\sim Ga(4.24, 1)$ and $\beta_4\sim G(1, 40)$. Finally, our error term variance parameter $\sigma$ was given a $Ga(1, 10)$ prior.

 \begin{figure}
\centering
\includegraphics[width=7cm,height=7cm]{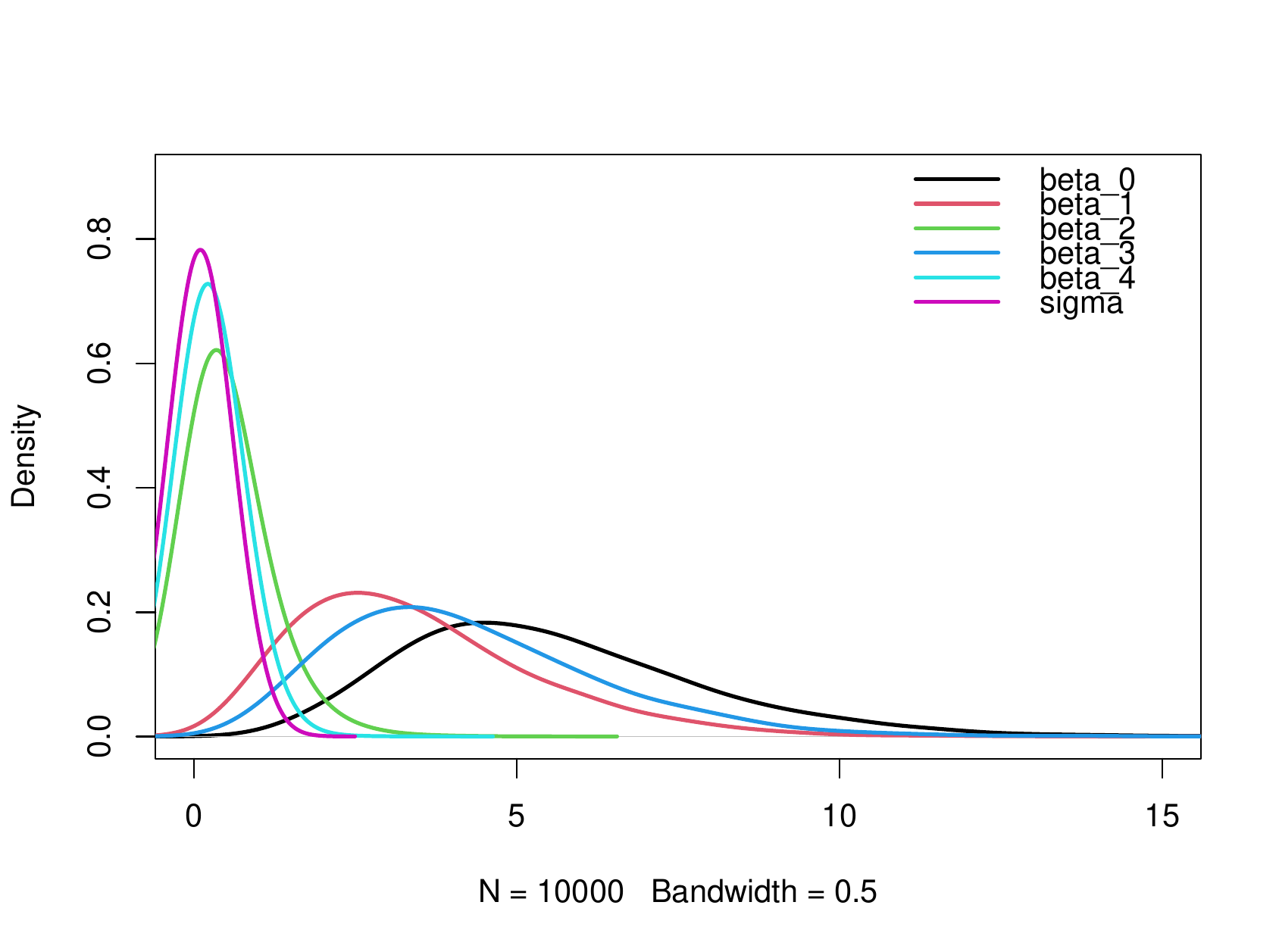}
\caption{Illustration of prior distribution for the $\beta_i$ parameters, $i=0,\dots,4$ and $\sigma$.}
\end{figure}
\
To sample the posterior distributions of these parameters, we used Hamiltonian Monte Carlo-NUTS using RStan \cite{team2016rstan}. For the Hamiltonian Monte Carlo-NUTS, we perform 10000 iterations with four different chains,  and discarding 5000 burn-in samples for each chain. The sampling took $29.731s$ in total. 

In Table ~\ref{t:bayescont}, we present the summary of these posterior simulations, giving the mean values for the fitted coefficients with $2.5\%$, and $97.5\%$ credible intervals from the Bayesian model fitting for Spring Barley, Winter Barley and Silage simulations. The nitrogen coefficients for all crops indicate a positive response to nitrogen to yield, but the phosphorus coefficient $\beta_4$ estimate is close to zero for all crops indicating a negligible response to $P$. We also tabulate the effective sample size, $n_{eff}$, and does vary considerably, and is generally $>25$\% of the number of samples for the parameter posterior distributions for all of our crops. 

\begin{figure}
\centering
\includegraphics[width=12cm,height=9cm]{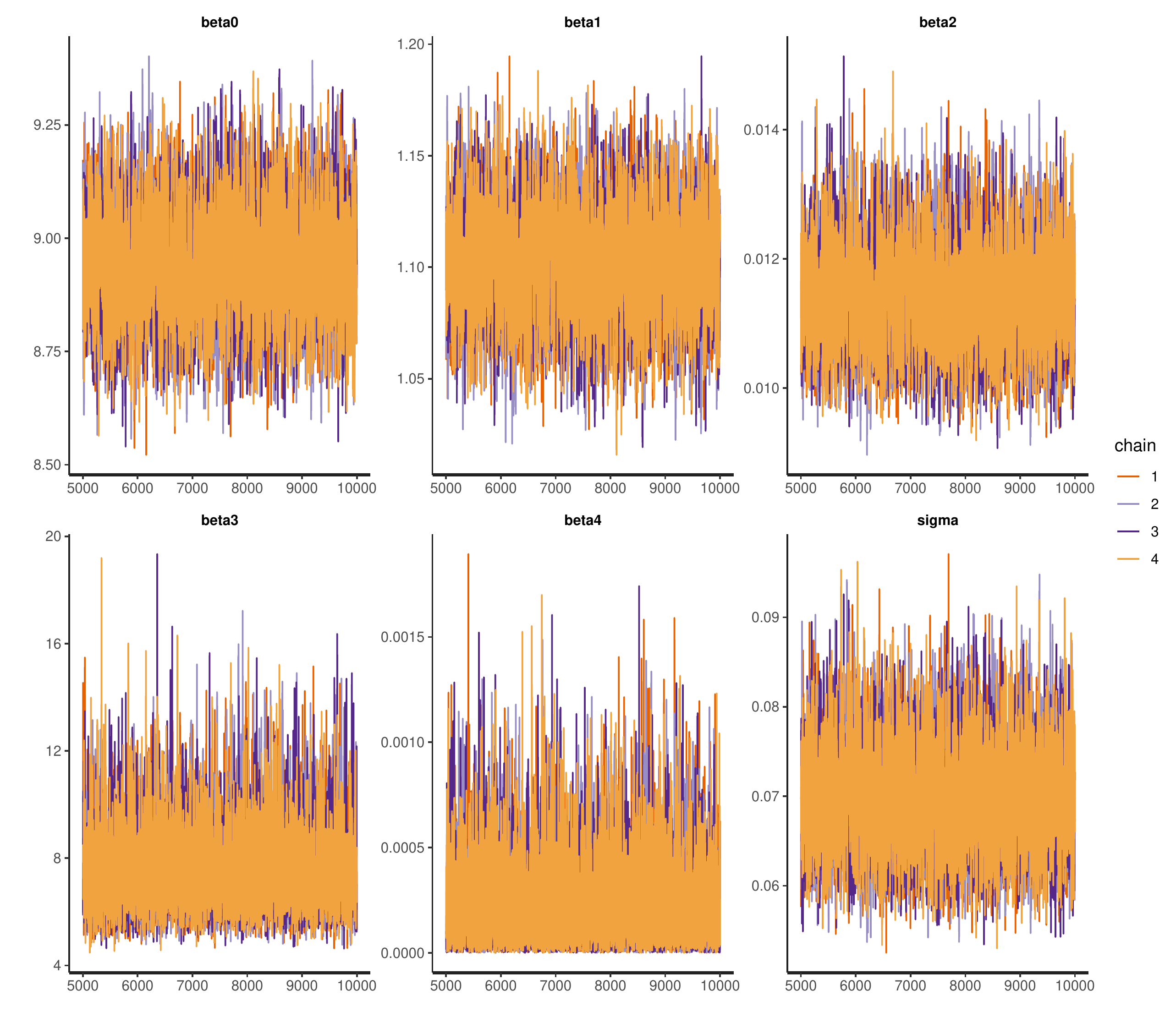}
\caption{Trace plot for the crop Winter Barley.}
\label{fig:7}
\end{figure}

We also evaluate $\hat{R}$, the potential scale reduction factor, also known as Gelman-Rubin statistic \citep{gelman2013bayesian} to summarise each chain in the sampler.  This statistic has the property that values  between $1.00$ to $1.01$ indicate that our chains are largely indistinguishable from one another, suggesting no evidence of lack of convergence. For the chains corresponding to the results in ~\ref{t:bayescont}, all the $\hat{R}$ values are lying between $1.00$ to $1.01$ suggesting a suitable level of convergence had been achieved. Diagnostic trace plots for the four Winter Barley chains are shown in Figure ~\ref{fig:7} and corroborate this, indicating generally good mixing and no signs of lack of convergence of the chains. Other diagnostics and plots for other crops are presented in the Appendix, but showed no features of concern.

\begin{table}%%[H]
\small
\caption{Posterior sample summary statistics $(1.0<\hat{R}<1.01)$ }
\label{t:bayescont}
\begin{center}
\begin{tabular}{|c|c|c|c|c|c|c|}
\hline
\hline
Crops& Coefficients& Mean & 2.5\% & 97.5\% &$n_{eff}$   \\
\hline
&$\beta_0$&5.005&4.775&5.345&4914\\
&$\beta_1$&0.4778&0.450&0.509&7882\\
Spring Barley &$\beta_2$&0.019&0.011&0.023&5360\\
&$\beta_3$&7.610&5.354&11.420&8842\\
&$\beta_4$&0.00009&0.000&0.0003&16927\\
&$\sigma$&0.17&0.140&0.220&5517\\
\hline
\hline
&$\beta_0$&8.948&8.719&9.296&4232\\
&$\beta_1$&1.102&1.055&1.145&4263\\
Winter Barley &$\beta_2$&0.011&0.009&0.013&4190\\
&$\beta_3$&7.617&5.428&11.231&8617\\
&$\beta_4$&0.0002&0.000&0.0007&14695\\
&$\sigma$&0.071&0.059&0.082&5128\\
\hline
\hline
&$\beta_0$&16.271&15.057&17.409&6103\\
&$\beta_1$&0.838&0.724&1.340&7205\\
Silage &$\beta_2$&0.016&0.012&0.17&6437\\
&$\beta_3$&13.03&7.036&20.917&13999\\
&$\beta_4$&0.005&0.0001&0.018&17489\\
&$\sigma$&0.127&0.011&0.143&12993\\
\hline
\end{tabular}
\end{center}
\end{table}

\begin{figure}%%[H]
     \centering
     \begin{subfigure}[b]{0.32\textwidth}
         \centering
         \includegraphics[width=\textwidth]{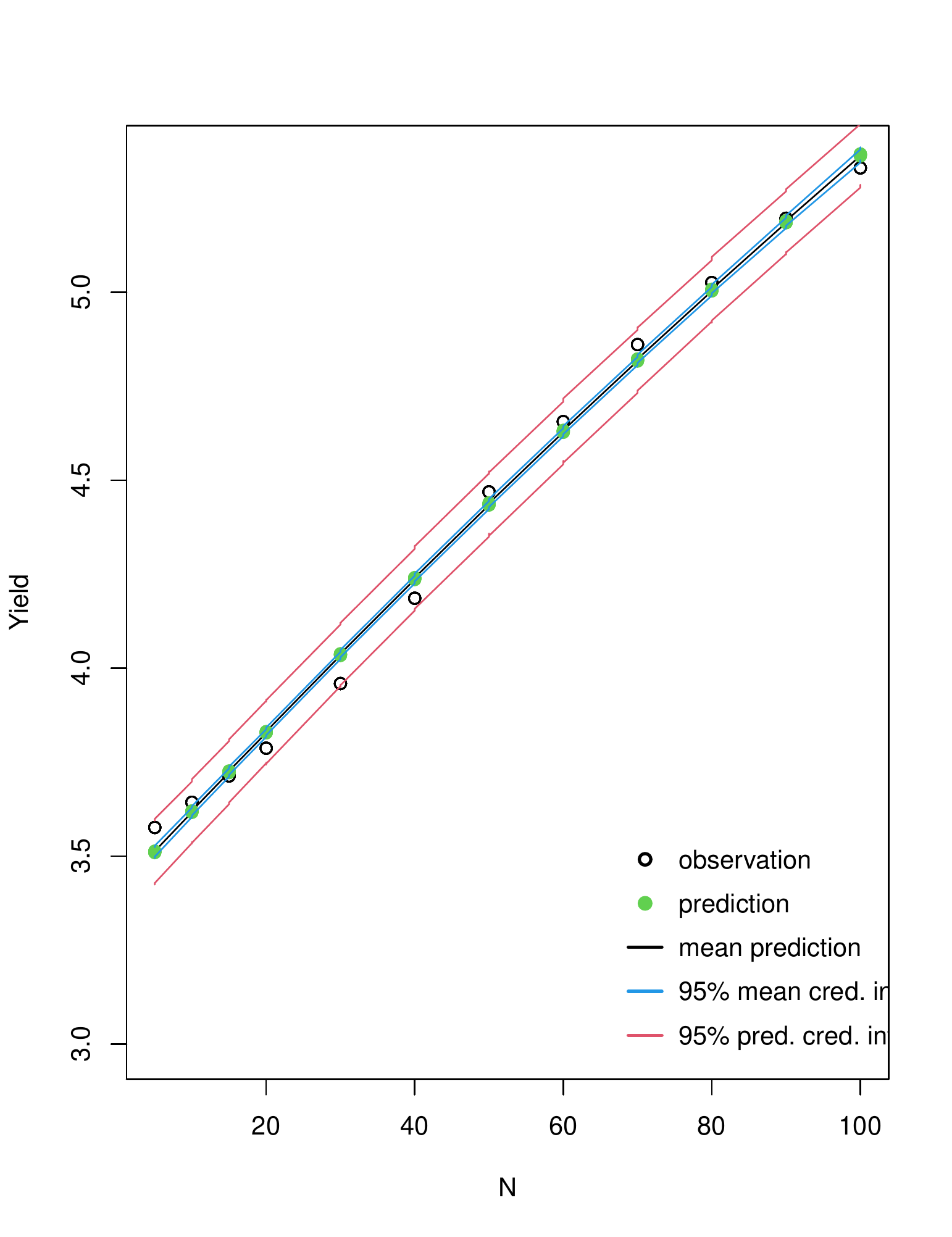}
     \end{subfigure}
     \hfill
     \begin{subfigure}[b]{0.32\textwidth}
         \centering
         \includegraphics[width=\textwidth]{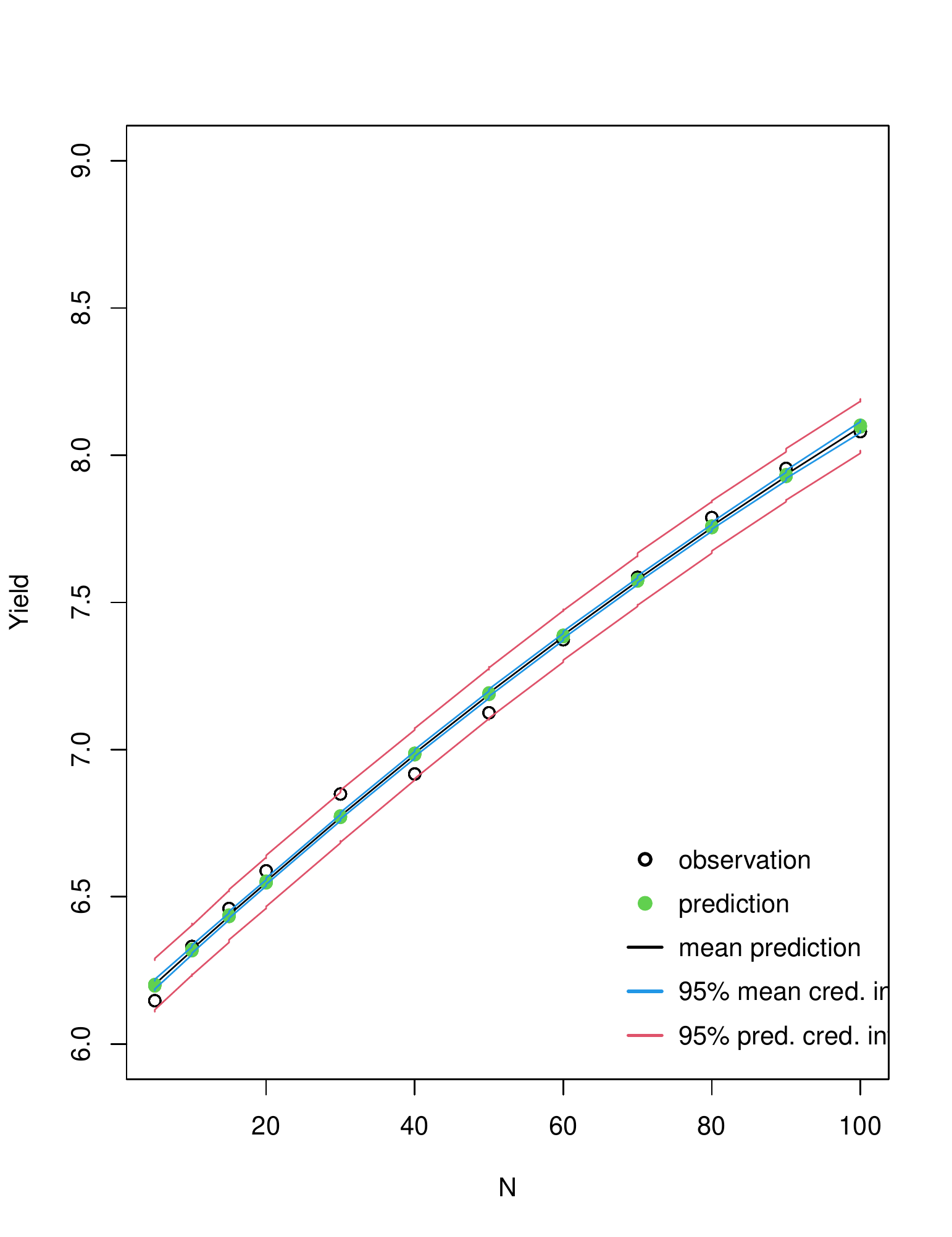}
     \end{subfigure}
      \hfill
     \begin{subfigure}[b]{0.32\textwidth}
         \centering
         \includegraphics[width=\textwidth]{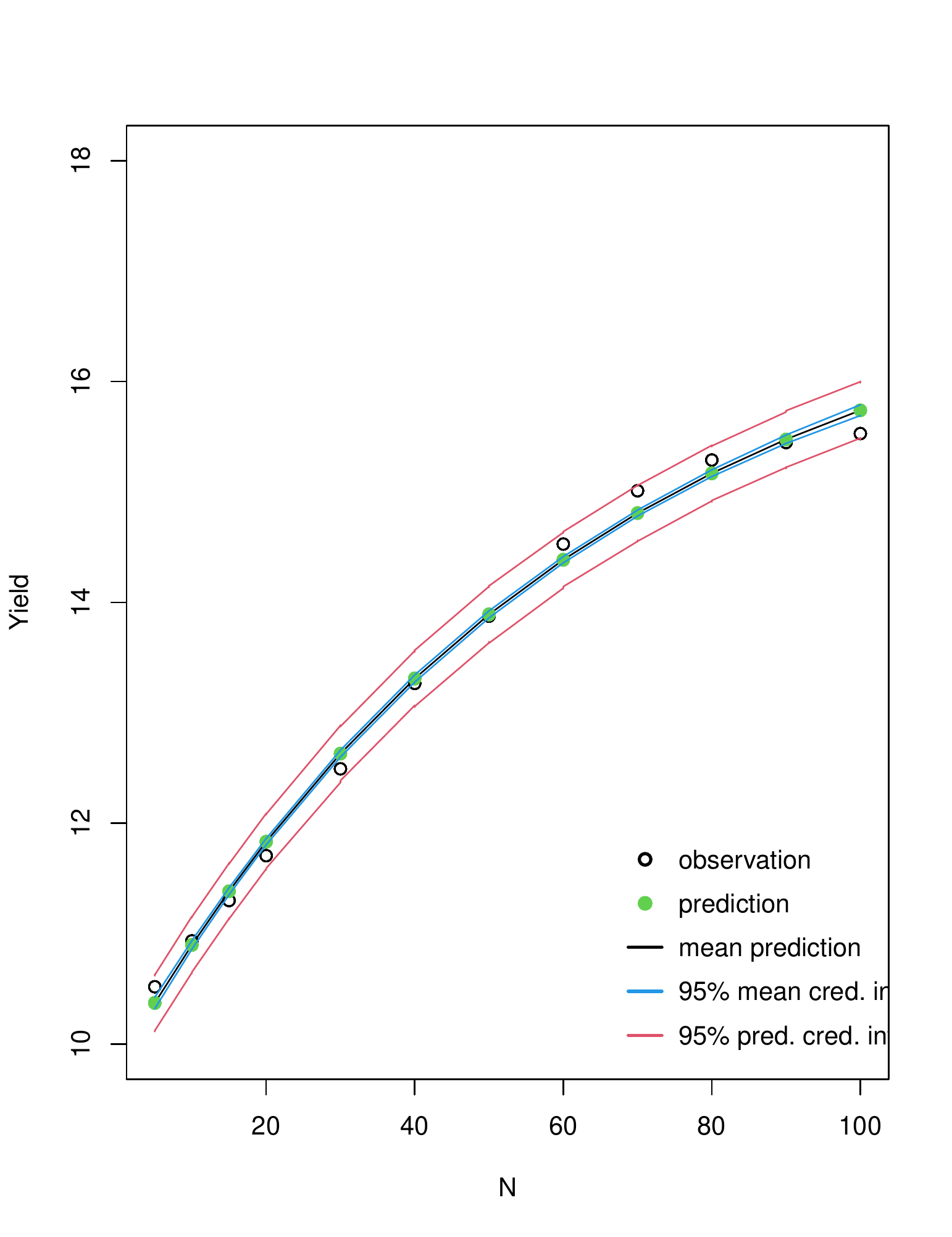}
     \end{subfigure}
   \caption{Non-linear Bayesian model fitting for spring barley (right); winter barley (middle); silage (right).}  
  \label{fig:6}    
\end{figure}

Synthesising these results, in Figure ~\ref{fig:6}, we plot the yield data (black circles) along side the point predictions (green circles), predicted yield curve (black line), 95\% mean credible intervals (blue line) indicates the credible intervals corresponding to the mean function of the Bayesian structure, and 95\% predicted credible intervals (red line) indicates the credible intervals corresponding to the mean function and $\sigma$ of the Bayesian structure. We note that the simulated data
lie within the mean and predicted credible intervals with very narrower uncertainty.

%%\vspace{-20pt}
\subsection{Model comparison and validation}
%%\vspace{-10pt}
Given the results observed above, where the influence of Phosphorous in the model appeared negligible, we now move to the question of model comparison. Clearly it is reasonable to consider whether it is meaningful to retain Phosphorous in the yield model after observing such weak dependency. We consider three models corresponding to a nitrogen-only response, a phosphorus-only response, and both Nitrogen and phosphorous responses. To select between the models, we apply the criteria of Subsection ~\ref{3.3}. Evaluating the ELPD, LOOIC, and WAIC statistics, and the Bayes factors give the results in Table ~\ref{t:comparison}, where the favoured model is highlighted in red. The Bayes factors are computed against the $N$-only model as the null model, with the $P$-only, and $N+P$ models as the alternatives.

From Table ~\ref{t:comparison}, we see that for all of our crops we favour the $N$-only model. We observe that the values of ELPD are maximised for the $N$-only model, and minimised for the  LOOIC and WAIC as desired. The Bayes factors of the alternative models are almost zero, indicating strong evidence against those alternatives. 

  \begin{table}%%[H]
\caption{Bayesian model comparison results.} 
\label{t:comparison}
\begin{center}
\begin{tabular}{|c|c|c|c|c|c|c|}
\hline
 Crops& Models &ELPD &LOOIC&WAIC &BF  \\
 \hline
&$N$&\color{red}{251.1}&\color{red}{-502.2}&\color{red}{-502.3}& \\

{Spring Barley}&$P$&-135.6&271.1&271&0.00\\

&$N+P$&250.8&-501.7&-501.3&0.0001\\
\hline
&$N$&\color{red}246.1&\color{red}-492.2&\color{red}-492.3&\\

{Winter Barley}&$P$&-140.6&281.1&281.1&0.00\\

&$N+P$&245.7&-491.3&-491.3&0.01\\
\hline

&$N$&\color{red}91.2&\color{red}-182.6&\color{red}-182.34&\\

{Silage}&$P$&-290.9&581.7&581.71&0.00\\

&$N+P$&91.1&-182.6&-182.16&0.00\\
\hline
\end{tabular}
\end{center}
\end{table}

\subsection{Incorporation of a factor variable}
In our data set, we have three different factor inputs:  steepness, soil, and weather. Adopting the approach described in ~\ref{sec:4} and considering an $N$-only model as suggested by the model selection above, we consider the inclusion of categorical variables into the model for winter barley.

For this analysis, we consider the effect of including a single factor variable into the model, as in \ref{sec:4}. We note that the three factor variables are such that we have four levels of steepness, five levels of soil, and six levels of weather. In each model, we take the first factor level as the baseline value giving up to five additional parameters denoted by the vector $\gamma_1$ representing the deviation of the maximum yield from this baseline value. Absent any particular information on the behaviour of the model under the different factor levels, independent broad Normal priors were adopted for each component of  $\gamma_1$ with mean $0$ and variance $1$.

Applying the same Hamiltonian Monte Carlo-NUTS approach used previously, we obtained the summary statistics given in Table ~\ref{t:factors} from our posterior simulations. The introduction of the factors to the model have increased the overall acceptance rate, and we also note that the posterior mean baseline maximum yield, $\gamma_0$, has increased compared to $\beta_0$ estimate obtained previously in Table \ref{t:bayescontN}.

In terms of diagnostics, all of the $\hat{R}$ values lie between $1.00$ to $1.01$ suggesting no evidence of lack of convergence of our posterior samples, and the effective sample sizes have improved significantly from the previous results. The trace plot of the model fitting for the factor steepness, soil, and weather shows good mixing of the chain for all the coefficients. Additionally, inspection of auto-correlation plots showed not many information is loosing because of thinning and posterior density plots of the factors also shows the normality.

In general, most of the estimates for the effects of different levels of the factors are close to zero or negative indicating substantial departures from the baseline maximum yield level. Additionally, we observe that the posterior estimates of the error variance parameter has decreased from $0.683$ substantially to values less than $0.23$ after adding the factors into our modelling, indicating narrower uncertainty due to resolving additional data variation through the expanded model. The estimated value of the coefficients $\beta_1$ and $\beta_2$ are remain same with high acceptance posterior samples after factoring compared to base result.

\begin{figure}
 \centering
\includegraphics[width=15cm]{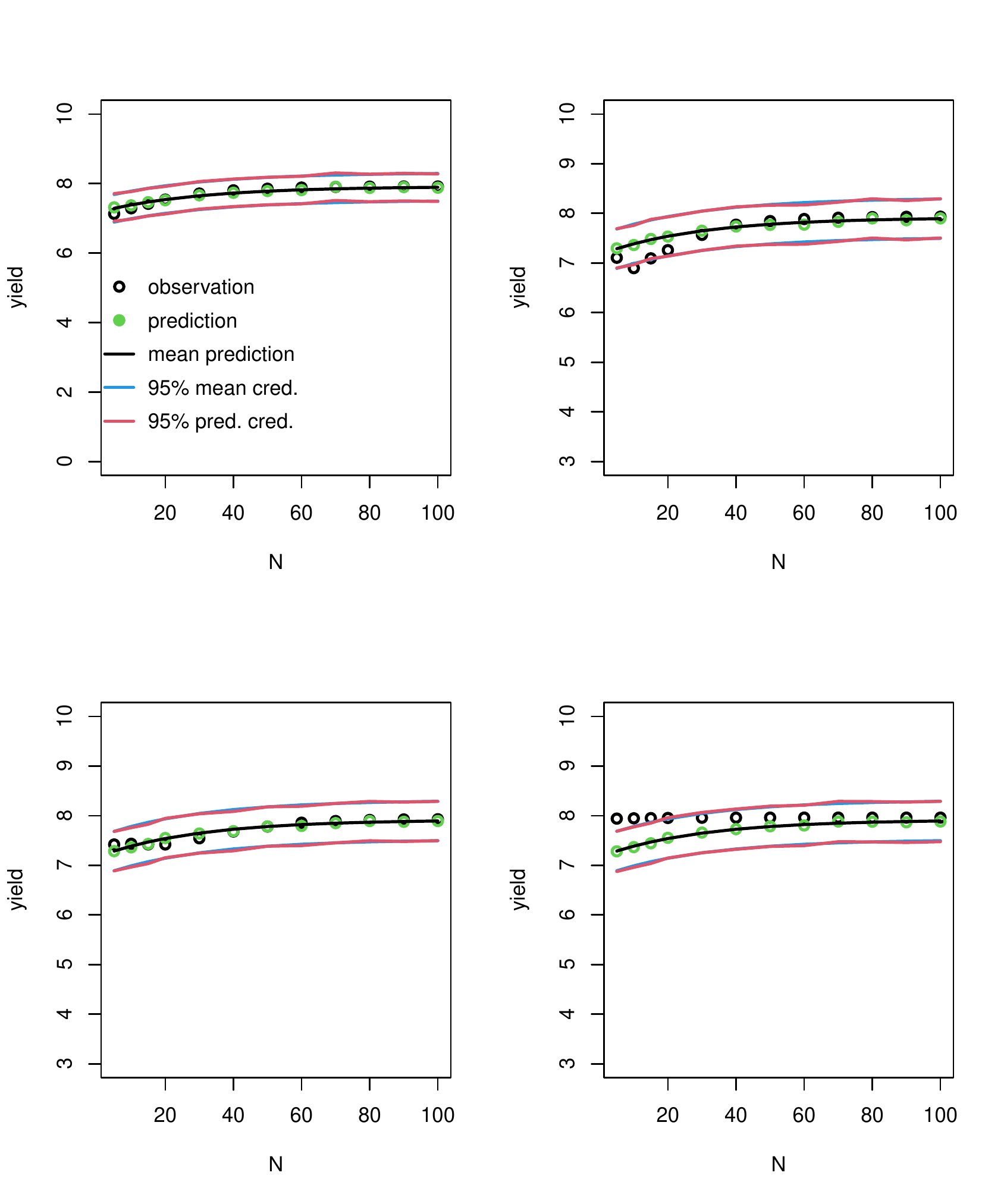}
\caption{ Upper panel: Prediction plot for the factor Steepness baseline (left) and level 1 (right); Lower panel: Prediction plot for the factor Steepness level 3 (left) and level 4 (right)}
 \label{fig:8(a)}
\end{figure}

Generating plots of the predicted yield as a function of $N$ in Figure ~\ref{fig:8(a)} including the steepness factor, we see that while the baseline and level 3 plots (left) yield good and consistent predictions the results do not directly transfer to levels 2 and 3 (right). In these cases, we observe that the observed yield data fundamentally deviates from the expected behaviour in ways that either cannot be captured by the Mitscherlich-Baule curve at all (the local non-monotonicity for small values of $N$ in level 2), or by inclusion of a single factor effect affecting maximum yield alone (the flattening of the yield response at level 4). In either case, an improvement could be made by introducing a further factor effect to modify the $\beta_1$ parameter to permit different strengths of relationship between yield and $N$ at the different categorical levels.

 \begin{figure}
 \centering
\includegraphics[width=12cm]{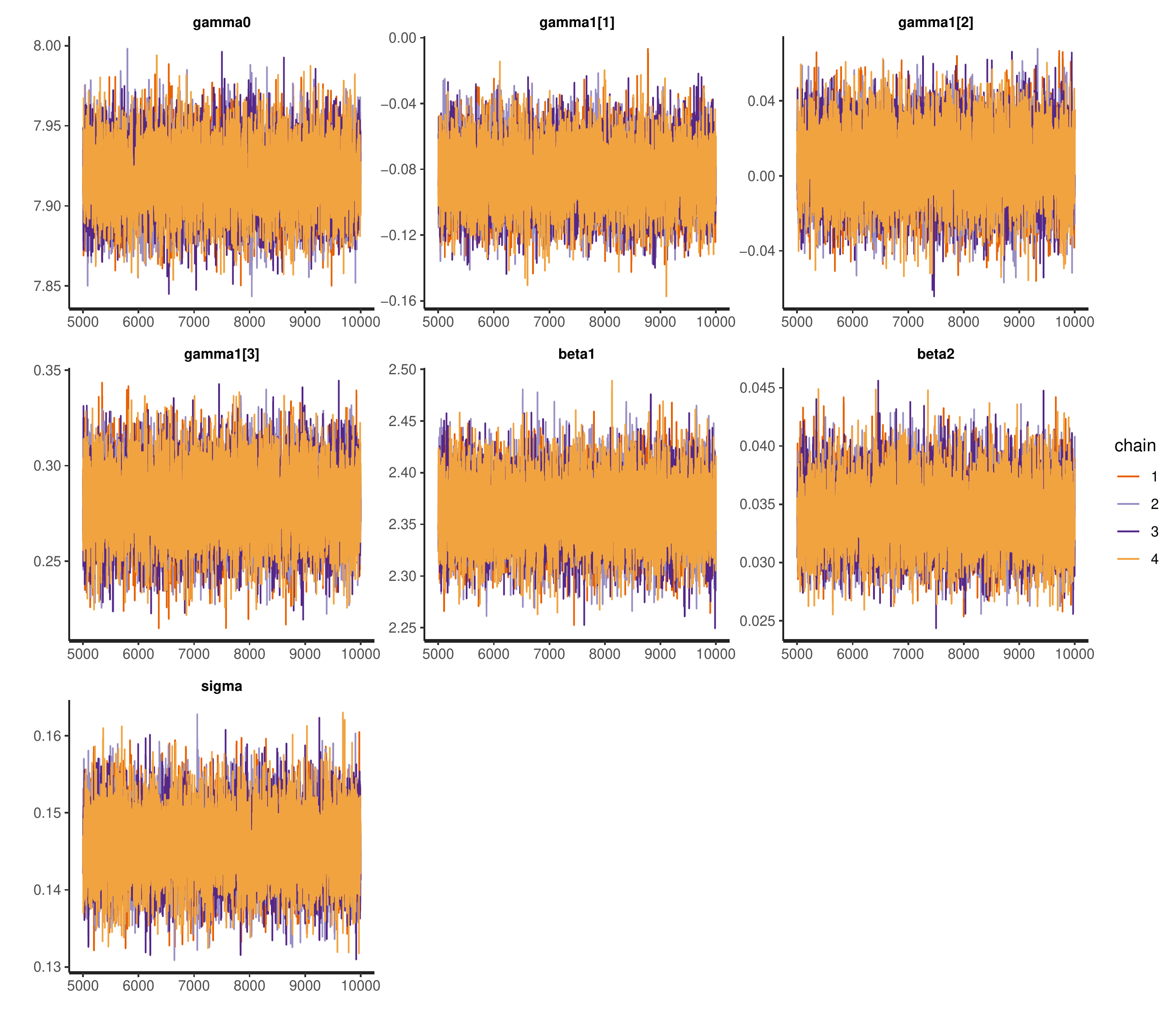}
\caption{Trace plot for the factor Steepness }
 \label{fig:8}
\end{figure}

\begin{table}%%[H]
\small
\caption{Posterior sample summary statistics for Winter Barley $N$ only model $(1.0<\hat{R}<1.01)$ }
\label{t:bayescontN}
\begin{center}
\begin{tabular}{|c|c|c|c|c|c|c|}
\hline
\hline
Crops& Coefficients& Mean & 2.5\% & 97.5\% &$n_{eff}$   \\
\hline
&$\beta_0$&6.945&6.873&7.019&11486\\
&$\beta_1$&2.514&2.140&3.010&9259\\
Winter Barley &$\beta_2$&0.053&0.034&0.077&4190\\
&$\sigma$&0.683&0.0650&0.714&12509\\
\hline
\end{tabular}
\end{center}
\end{table}
\begin{table}
\caption{Posterior sample summary statistics for $N$-only Winter Barley model, each including a single factor input $(1.0<\hat{R}<1.01)$} 
\label{t:factors}
\centering
\begin{tabular}{|c|c|c|c|c|c|c|}
\hline
\hline
\footnotesize
Factor& Coefficients& Mean & 2.5\% & 97.5\% &$n_{eff}$  \\
 \hline
&$\gamma_0$&7.916&7.879&7.957&11847\\
&$\beta_1$&2.361&2.341&2.421&16363\\
Steepness &$\beta_2$&0.035&0.028&0.039&12855\\
&$\mathbf{\gamma_{1,1}}$&-0.083& -0.118& -0.048&10926\\
&$\mathbf{\gamma_{1,2}}$& 0.006&-0.028& 0.041&14181\\
&$\mathbf{\gamma_{1,3}}$&0.281& 0.245&  0.314&14162\\
&$\sigma$& 0.145&0.137&0.154& 19320\\
\hline
\hline
&$\gamma_0$&8.035& 7.963&8.072&12017\\
&$\beta_1$& 2.112& 2.046&2.182&12202\\
Soil &$\beta_2$&0.029&0.022&0.036&11157\\
&$\mathbf{\gamma_{1,1}}$&-1.071&-1.126&-1.015&8846\\
&$\mathbf{\gamma_{1,2}}$& -0.005& -0.058&0.048&9975\\
&$\mathbf{\gamma_{1,3}}$&-1.811&-1.867& -1.755&8154\\
&$\mathbf{\gamma_{1,4}}$&-1.250&-1.305&-1.195&8476\\
&$\sigma$& 0.223& 0.213&0.234&14628\\
 \hline
\hline
&$\gamma_0$&  7.894& 7.844&7.913&12900\\
&$\beta_1$& 2.512&2.434&2.595&14577\\
Weather &$\beta_2$&0.030&0.023&0.037&10187\\
&$\mathbf{\gamma_{1,1}}$&-1.055& -1.099&-1.011&8260\\
&$\mathbf{\gamma_{1,2}}$&-0.921&-0.965&-0.877&8381\\
&$\mathbf{\gamma_{1,3}}$&  -1.092&-1.136&-1.048&8138\\
&$\mathbf{\gamma_{1,4}}$&-0.958&-1.003&-0.914&8100\\
&$\mathbf{\gamma_{1,5}}$&-2.485&-2.532&-2.439&7940\\
&$\sigma$&0.184&0.176&  0.193&15481\\
\hline
\end{tabular}
\end{table}

\section{Concluding Remarks}
%%\vspace{-20pt}

A Bayesian hierarchical model has proven to be a useful and effective tool for modelling the behaviour of the key quantity of interest in our agricultural application. Basing an emulator around a known non-linear growth function allows us to embed known scientific knowledge directly into the model, while still affording some flexibility and variation due to the different conditions of the simulated experiments. The fully-Bayesian approach allowed for an appropriate modelling and capture of the uncertainties in the problem. However, without strong prior information to inform the process, we may prefer instead to seek an approach which is less computationally intensive but could deliver results of a similar quality. For example, a Bayes linear \cite{goldstein2007bayes} emulator would capture much for the uncertainty without requiring intensive simulation.

A key challenge that arises from the analysis of such large quantities of simulated output is that the scope, scale, and black-box nature of the synthetic experiments are such that it is inevitable that a non-trivial quantity of the simulator output will behave unexpectedly. The reasons for this could be simulator failure, numerical instability, or unexpectedly deleterious combinations of input parameters. Regardless of cause, a problematic outcome is that the growth curve at the heart of the emulator model is ill-equipped to model the simulated output under these circumstances. This may suggest that a more pragmatic approach would be to adopt a simpler mean function and allow for more of the structural behaviour of the response variable to be captures by the residual Gaussian process.

Finally, the presence of factor variables in a model such of this is a challenging topic for current research \cite{qian2008gaussian}. There is a spectrum of choices to be made about the role and presence of a categorical variable in the model’s mean function, even if that model is a simple polynomial. Additionally, when multiple factors with many levels are present this could lead to a profusion of additional parameters leading to challenges in estimation and identifiability. Furthermore, a more challenging question is to determine the appropriate way in which categorical variables enter the residual process and their impact on the correlation structure between the simulator evaluations.

\section*{Acknowledgements}
The authors are very grateful to Durham University for awarding a Durham Doctoral Scholarship to conduct this research. 

\section*{Declaration of conflicting interests}
The author declared no potential conflicts of interest with respect to the research, authorship and/or publication of this article.

\section*{Funding}
The author received no financial support for the research, authorship and/or publication of this article.

\small
\bibliography{smjtemplate}

\newpage
\appendix
 \begin{figure}
\centering
\includegraphics[width=15cm,height=6cm]{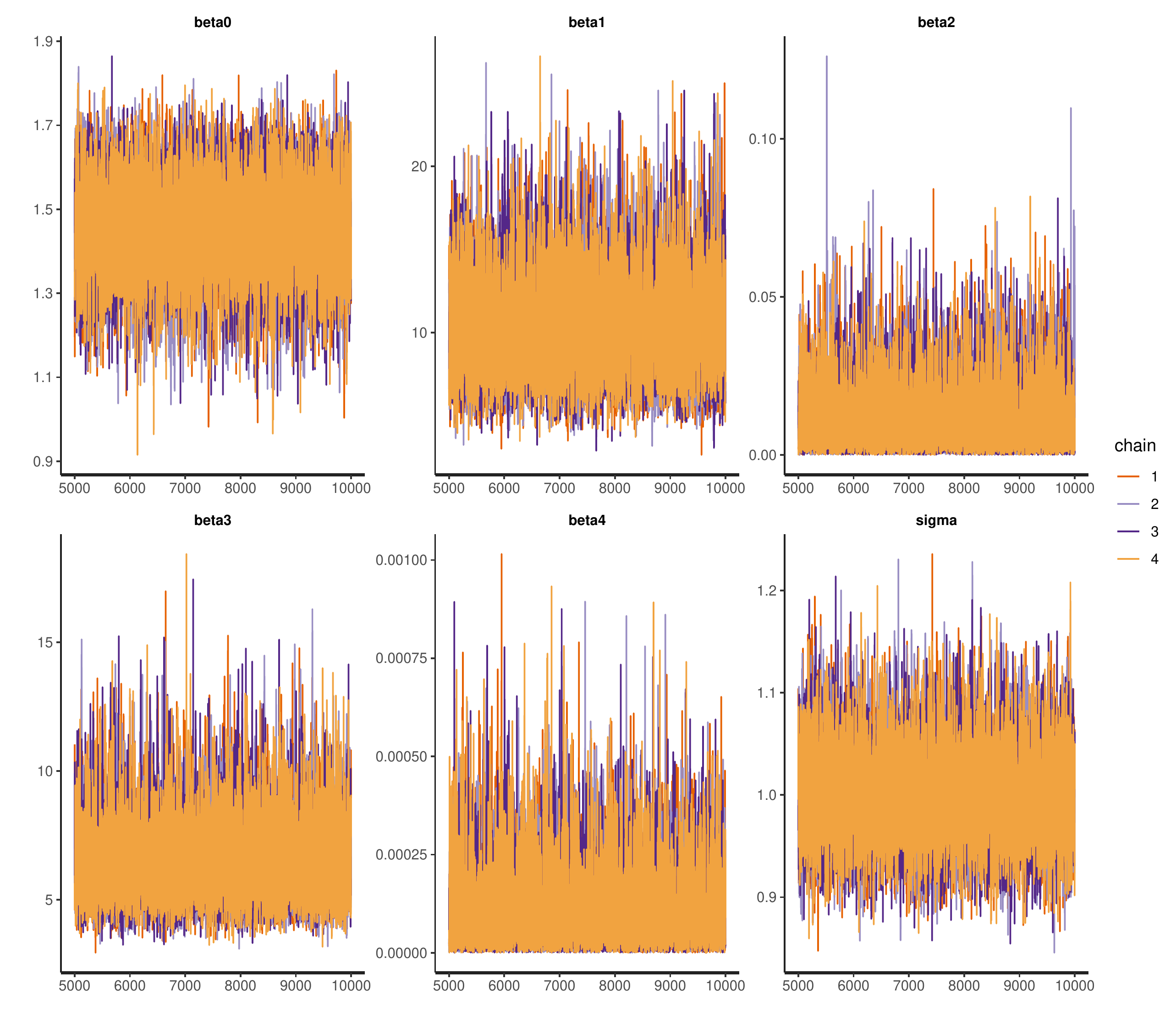}
\caption{Trace plot for the crop Spring Barley.}
\label{fig:11}
\end{figure}

 \begin{figure}
\centering
\includegraphics[width=15cm,height=6cm]{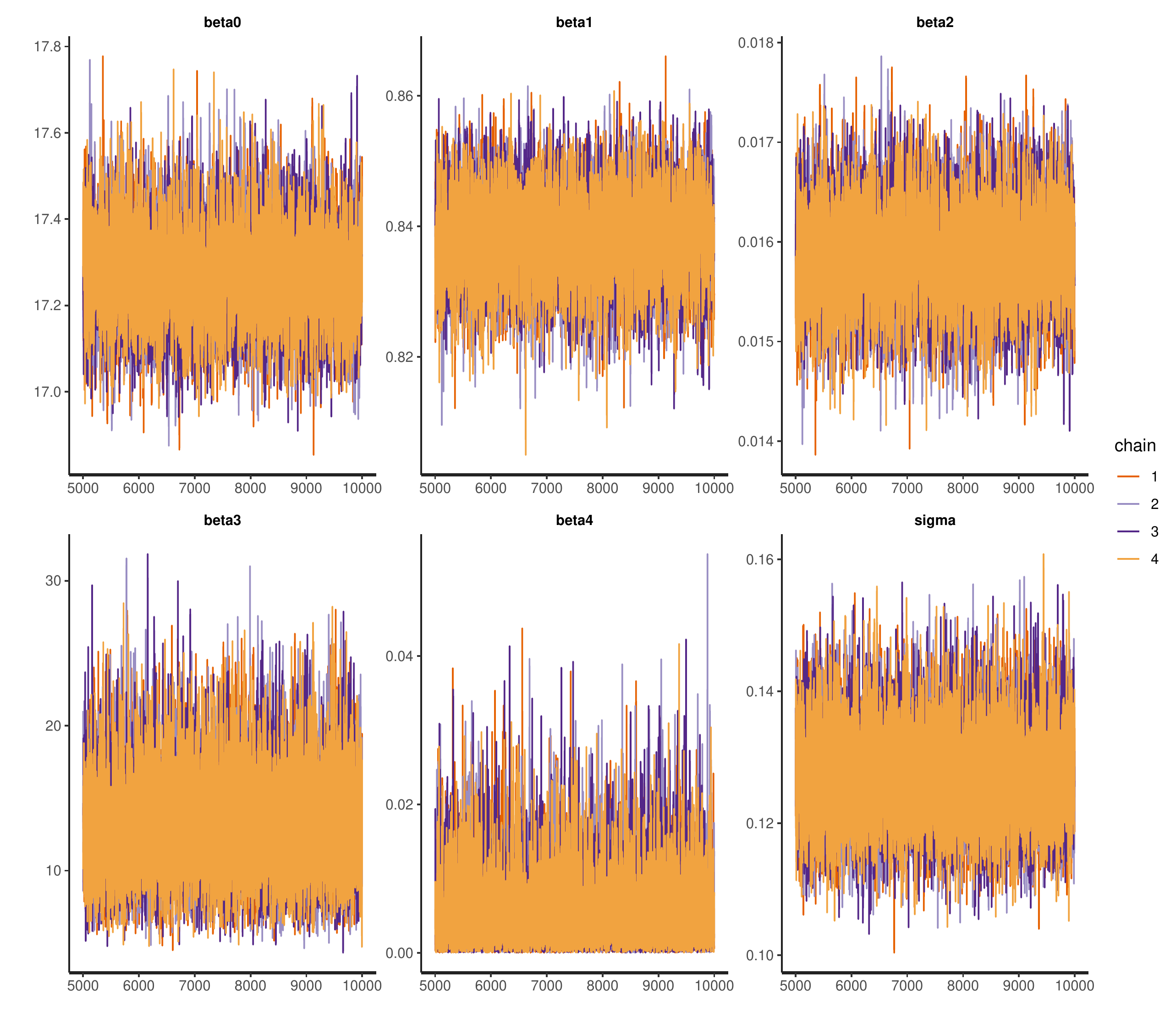}
\caption{Trace plot for the crop Silage.}
\end{figure}
 \begin{figure}
\centering
\includegraphics[width=15cm,height=6cm]{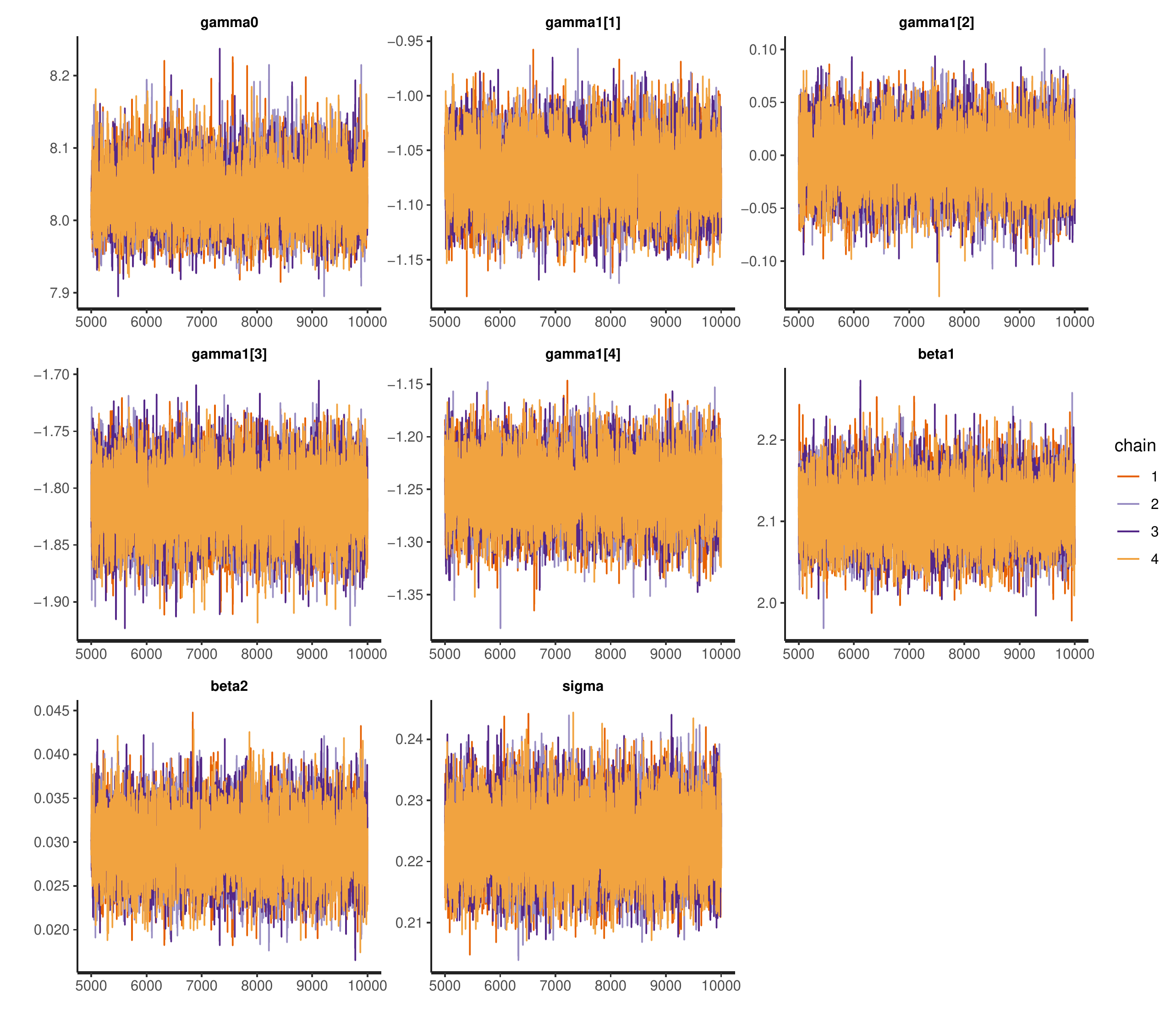}
\caption{Trace plot for the factor soil}
 \label{fig:9}
\end{figure}

 \begin{figure}
\centering
\includegraphics[width=17cm,height=8cm]{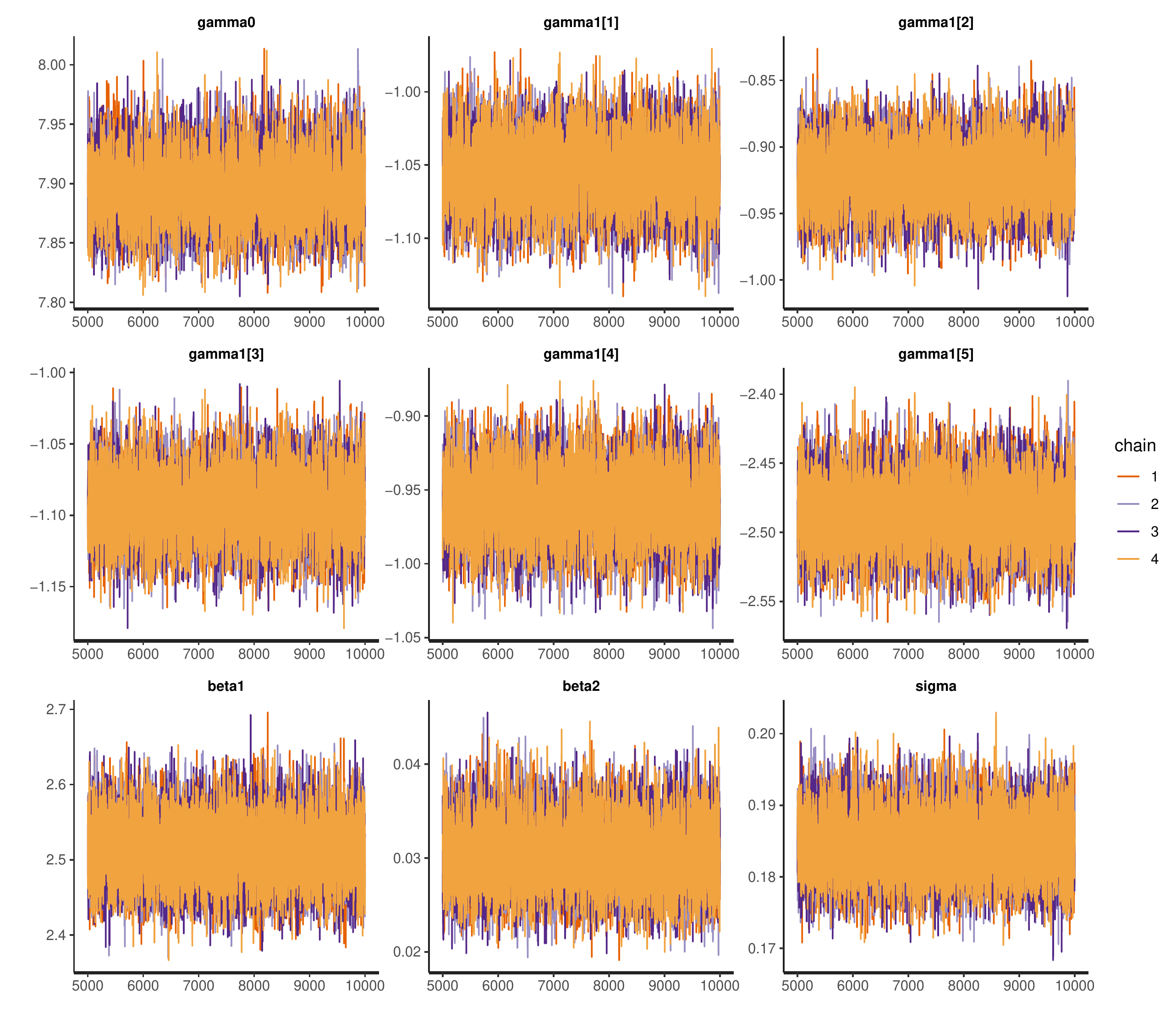}
\caption{Trace plot for the factor Weather}
 \label{fig:10}
\end{figure}

 \begin{figure}
\centering
\includegraphics[width=17cm,height=5cm]{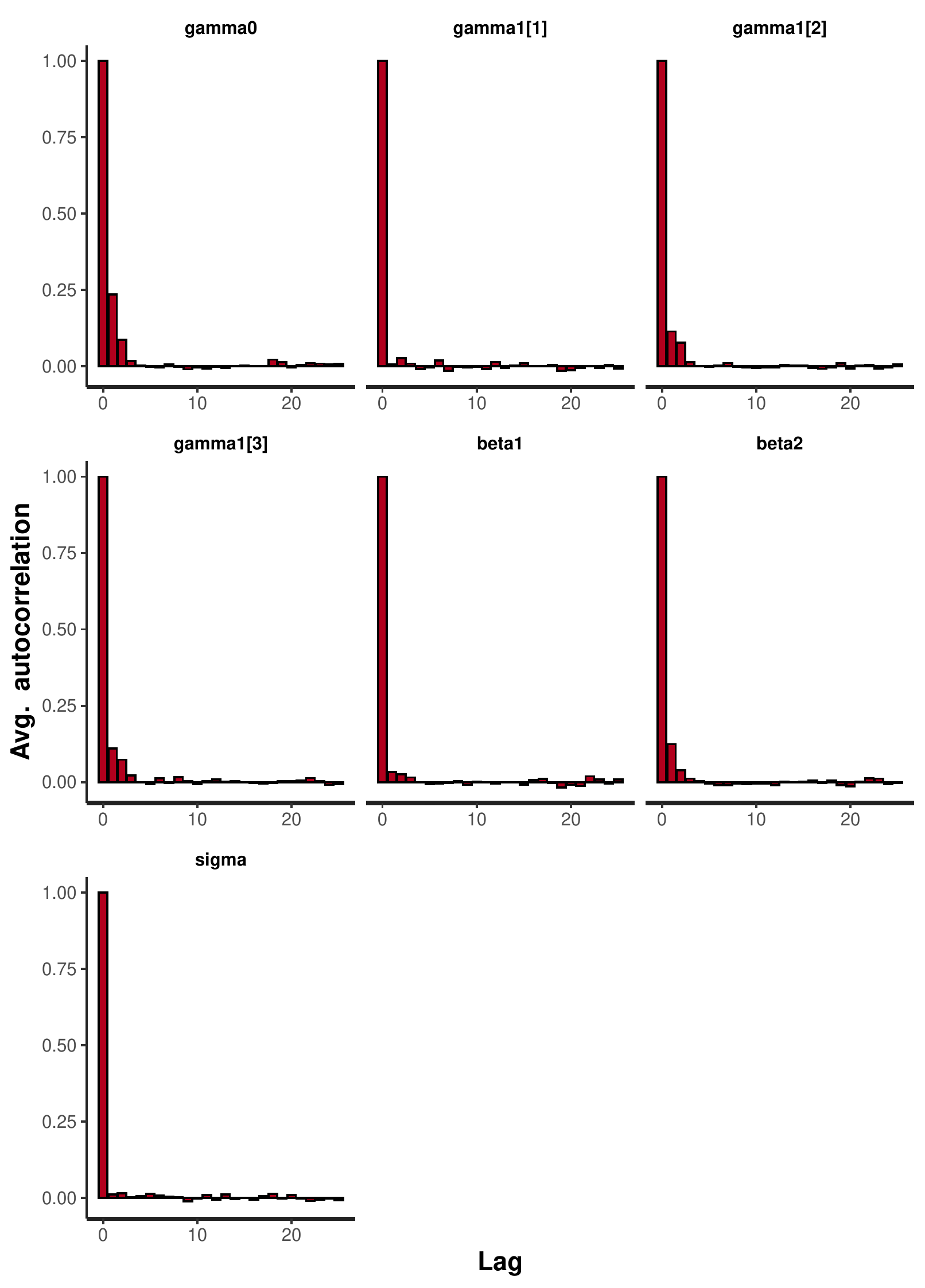}
\caption{Auto-correlation plot for the factor Steepness }
\label{fig:12}
\end{figure}
 \renewcommand{\thefigure}{14}
 \begin{figure}%[H]
\centering
\includegraphics[width=15cm,height=5cm]{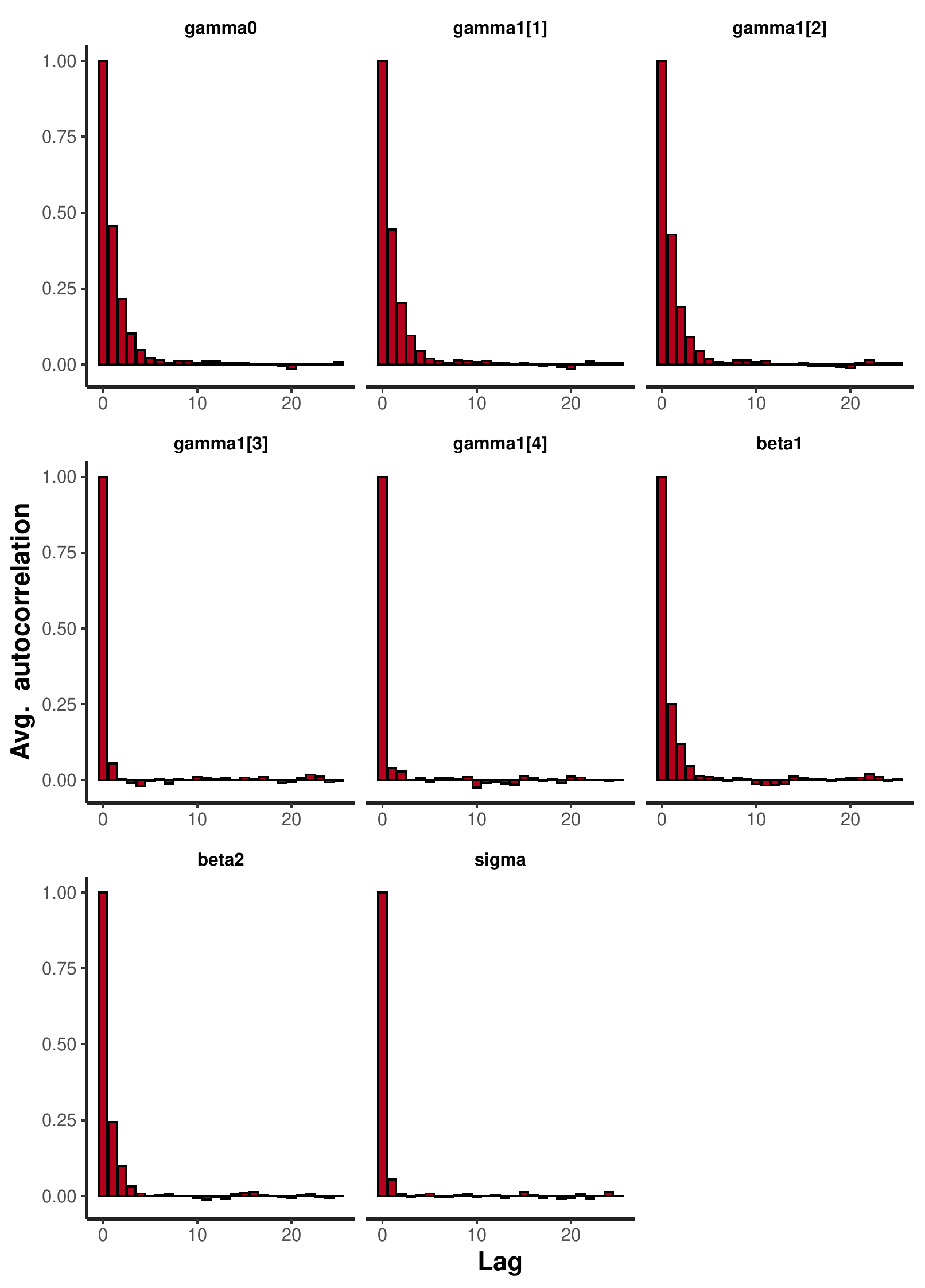}
\caption{Auto-correlation plot for the factor Soil}
\label{fig:13}
\end{figure}
 \renewcommand{\thefigure}{15}
\begin{figure}%[H]
\centering
\includegraphics[width=17cm,height=5cm]{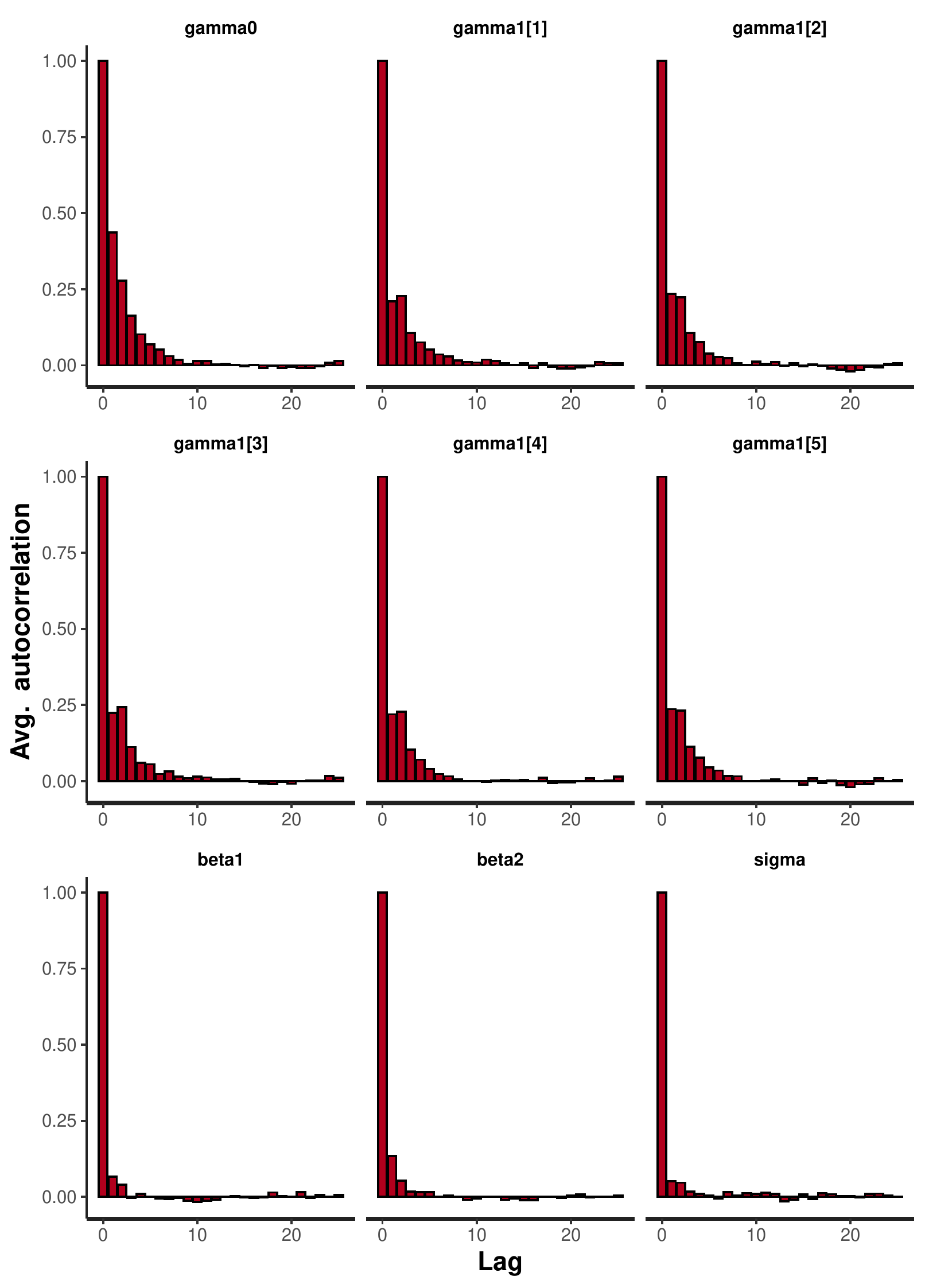}
\caption{Auto-correlation for the factor Weather}
\label{fig:14}
\end{figure}

\renewcommand{\thefigure}{16}
\begin{figure}%[H]
\centering
\includegraphics[width=17cm,height=5cm]{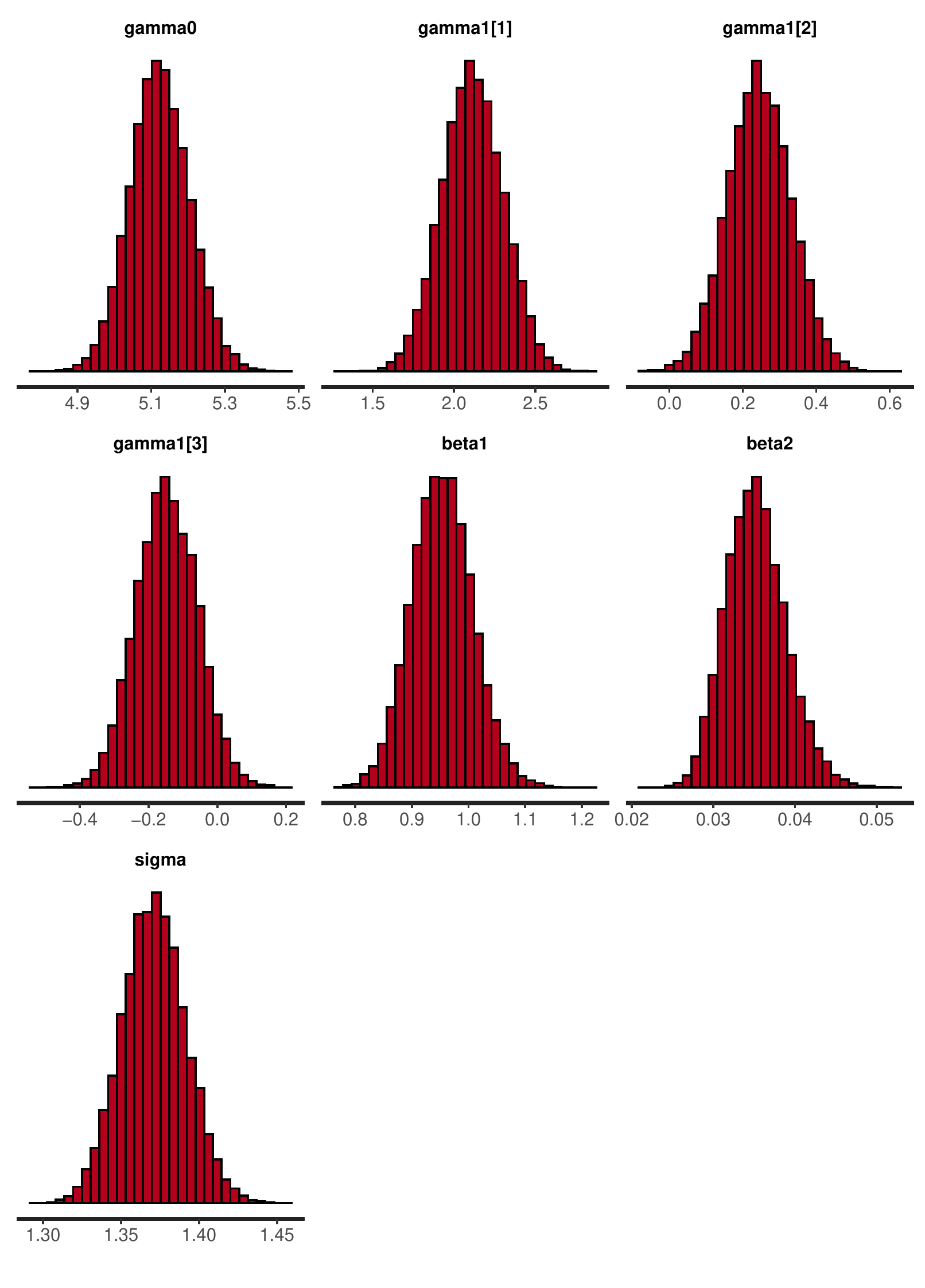}
\caption{Posterior density plot for the factor Steepness }
\label{fig:15}
\end{figure}

\renewcommand{\thefigure}{17}
\begin{figure}%[H]
\centering
\includegraphics[width=17cm,height=5cm]{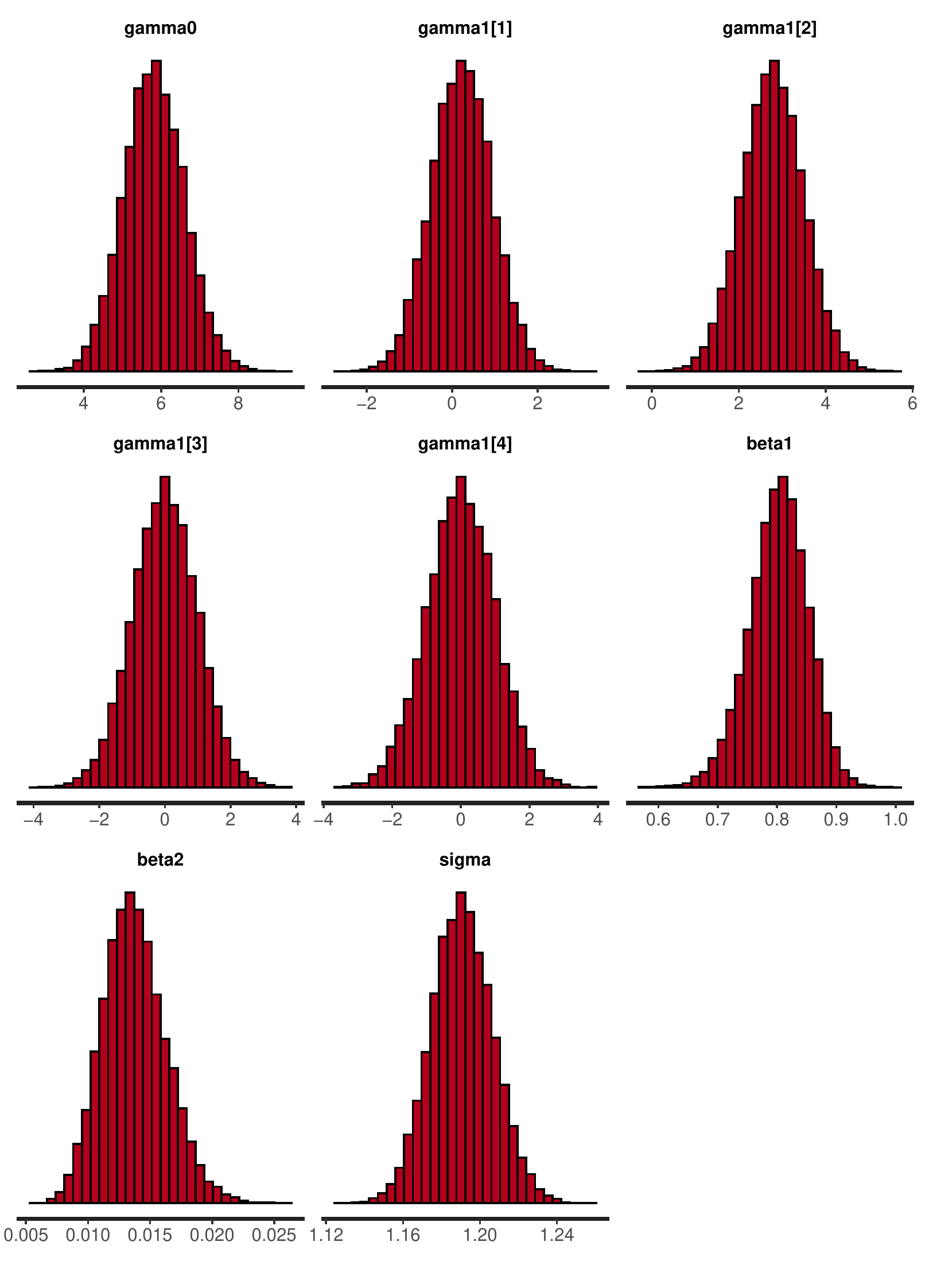}
\caption{Posterior density plot for the factor Soil }
\label{fig:16}
\end{figure}

\renewcommand{\thefigure}{18}
\begin{figure}%[H]
\centering
\includegraphics[width=17cm,height=5cm]{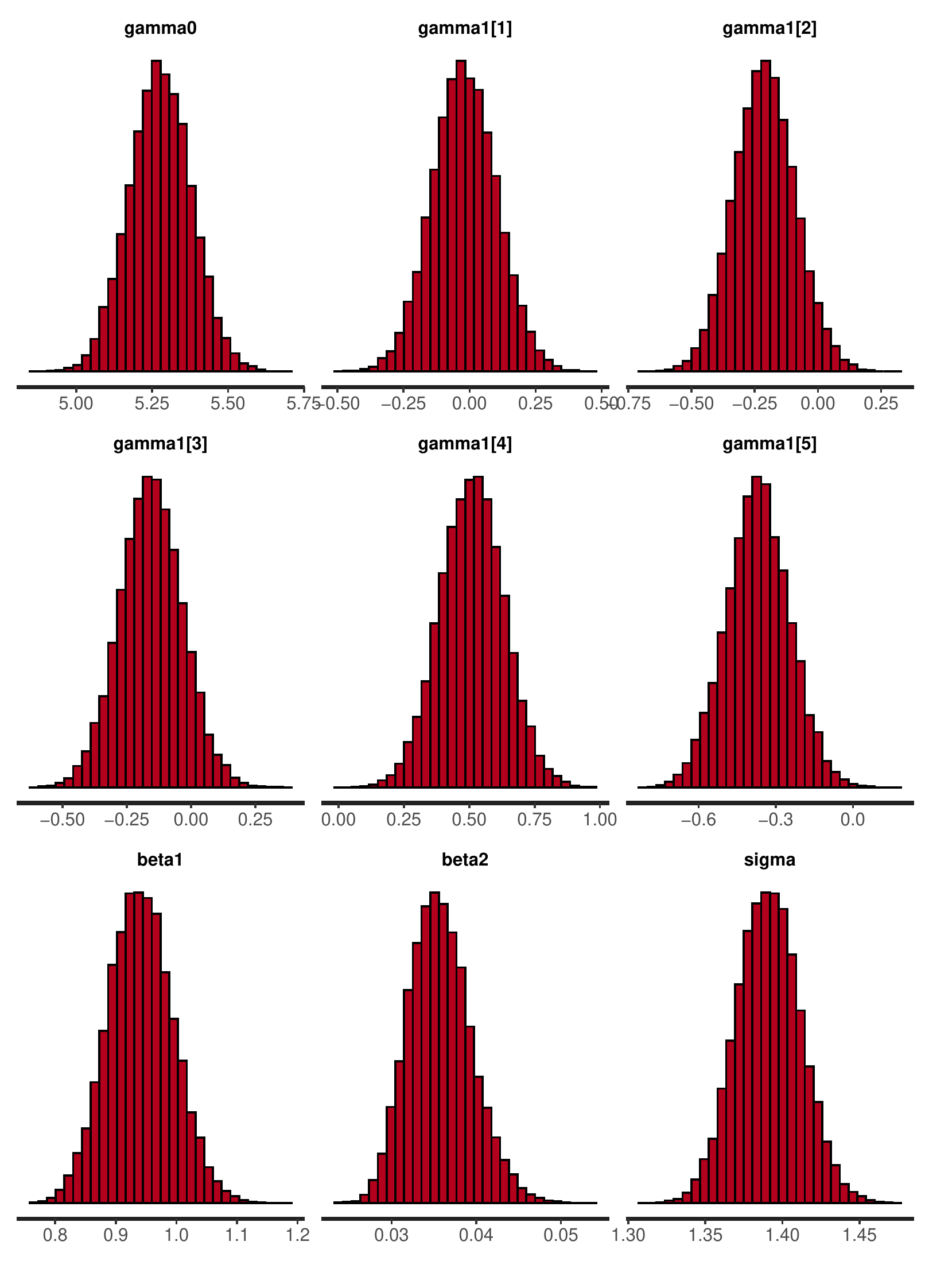}
\caption{Posterior density plot for the factor Weather}
\label{fig:17}
\end{figure}
\renewcommand{\thefigure}{19}
\begin{figure}
\centering
\includegraphics[width=12cm,height=9cm]{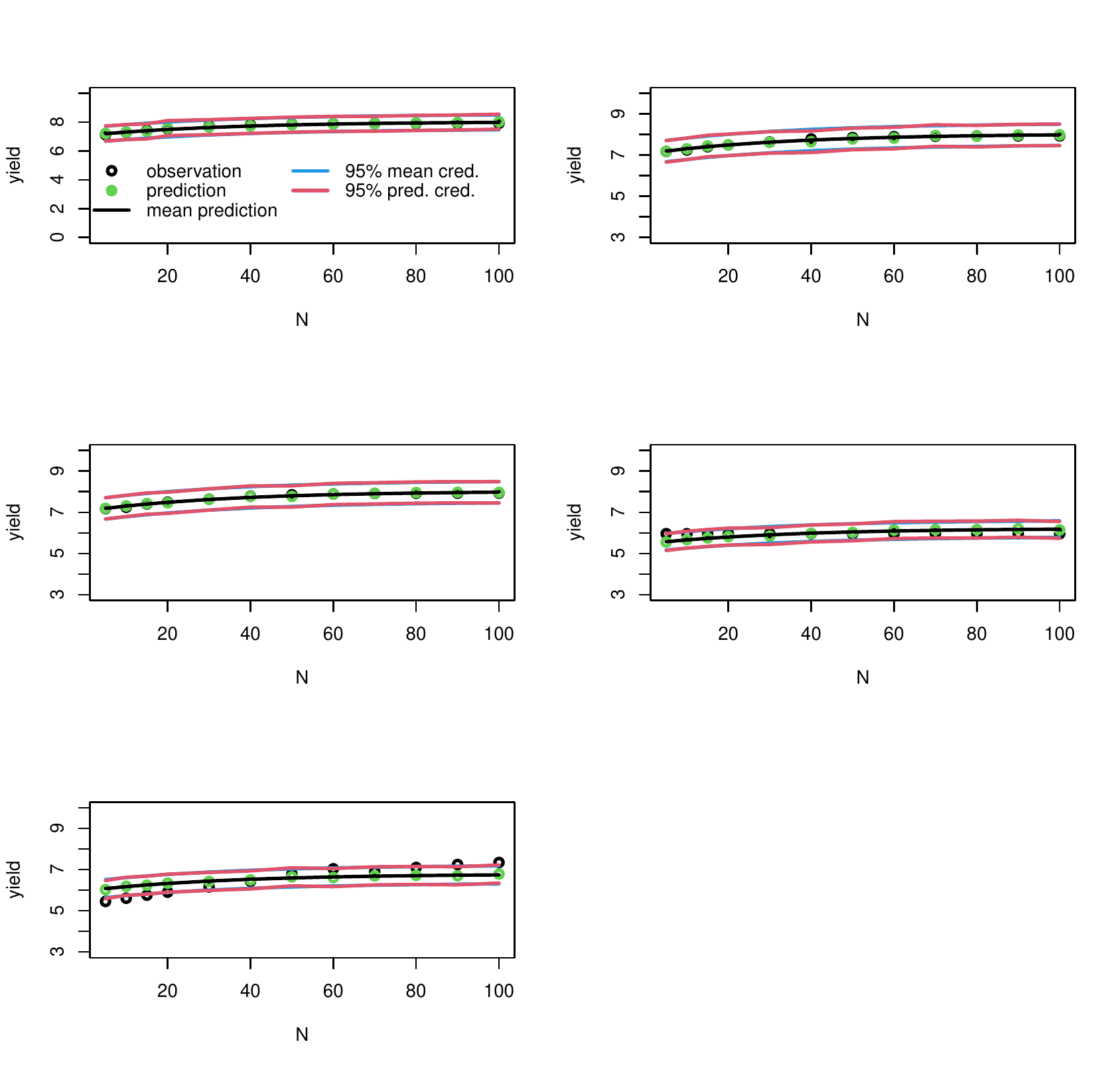}
\caption{Prediction plot for the factor Soil}
 \label{fig:8(B)}
\end{figure}
\renewcommand{\thefigure}{20}
\begin{figure}
\centering
\includegraphics[width=12cm,height=9cm]{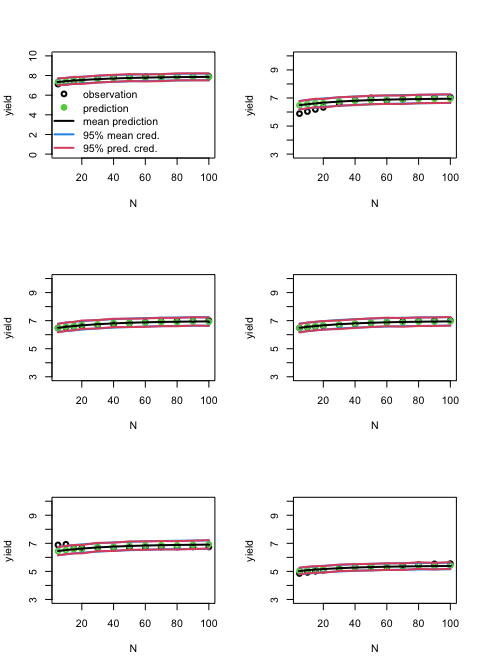}
\caption{Prediction plot for the factor Weather}
 \label{fig:8(c)}
\end{figure}

\end{document}